\documentclass[
 reprint,
 amsmath,
 pre,
 amssymb,
 nofootinbib,
 aps,
 longbibliography,
]{revtex4-1}

\usepackage[pdftex]{graphicx}
\usepackage{wasysym}
\usepackage{amsfonts}
\usepackage{ifsym}
\usepackage{pifont}
\usepackage{footmisc}
\usepackage{soul}

\makeatletter
\newlength{\apb@width}
\newcommand{\autoparbox}[2][c]{\settowidth{\apb@width}{#2}\parbox[#1]{\apb@width}{#2}}

\makeatother

\newcommand{\namedref}[2]{\hyperref[#2]{#1~\ref*{#2}}}

\newcommand{\tr}{\operatorname{tr}}
\newcommand{\Tr}{\operatorname{Tr}}

\newcommand{\Csphere}{{}^\bullet\kern-1.2pt C}
\newcommand{\Ctorus}{{}^\circ\kern-1.2pt C}

\newcommand{\nn}{\nonumber}

\newcommand{\COMMENT}[1]{}

\newcommand{\neqa}{\nonumber\end{eqnarray}}
\newcommand{\la}[1]{\label{#1}}

\newcommand{\<}{{\langle}}
\renewcommand{\>}{{\rangle}}

\newcommand{\re}{\relax{\rm I\kern-.18em R}}

\def\su2{{SU(2)}}

\def\[{\left[}
\def\]{\right]}

\def\({\left(}
\def\){\right)}
\def\[{\left[}
\def\]{\right]}

\def\<{\langle}
\def\>{\rangle}

\def\2F1{\,_2{\rm F}_1}

\usepackage[usenames,dvipsnames]{xcolor}
\usepackage{hyperref}
\hypersetup{
    colorlinks,
    linkcolor={red!60!black},
    citecolor={blue!70!black},
    urlcolor={blue!80!black}
}
\usepackage{mathbbol}

\usepackage{cancel}
\usepackage{tikz}
\usepackage{ctable}
\usepackage{booktabs}
\usepackage[caption=false]{subfig}
\newcolumntype{L}[1]{>{\raggedright\let\newline\\\arraybackslash\hspace{0pt}}m{#1}}
\newcolumntype{C}[1]{>{\centering\let\newline\\\arraybackslash\hspace{0pt}}m{#1}}
\newcolumntype{R}[1]{>{\raggedleft\let\newline\\\arraybackslash\hspace{0pt}}m{#1}}
\usepackage{mathtools}

\DeclarePairedDelimiter\ket{\lvert}{\rangle}
\DeclarePairedDelimiterX\braket[2]{\langle}{\rangle}{#1 \delimsize\vert #2}

\newcommand{\beq}{\begin{equation}}
\newcommand{\eeq}{\end{equation}}
\newcommand{\beqq}{\begin{equation*}}
\newcommand{\eeqq}{\end{equation*}}
\newcommand\beqa{\begin{eqnarray}}
\newcommand\eeqa{\end{eqnarray}}
\newcommand\beqaa{\begin{eqnarray*}}
\newcommand\eeqaa{\end{eqnarray*}}
\newcommand\bea{\begin{array}}
\newcommand\eea{\end{array}}

\begin{document}


\title{Following Black Hole States}

\author{Kasia Budzik$^{a,b}$, 
Harish Murali$^{a,b}$, Pedro Vieira$^{a,c}$ 
}
\affiliation{$^{a}$Perimeter Institute for Theoretical Physics, 31 Caroline St N, Waterloo, Ontario N2L 2Y5, Canada}
\affiliation{$^{b}$Department of Physics and Astronomy, University of Waterloo, Waterloo, Ontario, N2L 3G1, Canada}
\affiliation{$^{c}$Instituto de F\'isica Te\'orica, UNESP, ICTP South American Institute for Fundamental Research, Rua Dr Bento Teobaldo Ferraz 271, 01140-070, S\~ao Paulo, Brazil}


\begin{abstract}
\emph{How an old sage by the name of K ate two gravitons and was never seen again (because he became a black hole).} \\

We study $\mathcal{N}=4$ SYM at non-integer number of colours. By varying $N$ we can continuously follow states all the way from $N=\infty$ where integrability reigns to finite $N$ where quantum gravity effects dominate. As an application we consider classically $1/16$-BPS states. Quantum mechanically, these states are generically non-supersymmetric but some special states -- at special values of $N$ -- become super-symmetric at the quantum level as well. They are the so-called quantum black hole states studied recently using cohomology. We write down the form of the lightest BH state at $N=2$ -- and follow it in $N$, both at weak coupling and -- more speculatively -- at strong coupling as well. At weak coupling this state has protected dimension~$\Delta=19/2$ at $N=2$ and becomes a triple trace made out of Konishi and two light BPS operators at infinite~$N$ with~$\Delta=19/2+12\lambda+\dots$. At strong coupling we suspect it becomes a quadruple trace with dimension~$\Delta \simeq 19/2+\text{integer}$.
\end{abstract}


\pacs{Valid PACS appear here}
\maketitle

\section{Introduction} \label{sec:introduction}

Contemplating a black hole wave function is a whimsical dream which AdS/CFT translates into a very concrete task: Finding the eigenvectors of the dual conformal field theory dilatation operator with large eigenvalues, comparable to the central charge of the field theory. 

One can attack this problem by focusing on a special corner of $\mathcal{N}=4$ super Yang-Mills consisting of  operators which preserve $1/16$ of the supersymmetry. 
Ultimately, to study classical black holes (BH) with large energy and a small Newton constant we will have to consider operators with a parametrically large number of scalars and a large number of colours $N$. 
A warm-up problem is to find $1/16$-th BPS operators with a small number of constituents and a small number of colours.

The simplest kind of $1/16$-BPS states are called \textit{multigraviton} (MG) states and can be thought of as gases of gravitons. As the name suggests, they are given by products of gravitons, which are $\tfrac12$-BPS operators (or their descendants), plus $Q$-exact terms. It was observed in \cite{XiYinPaper, Grant:2008sk} that the number of MG states does not grow fast enough to account for the entropy of supersymmetric BHs~\cite{Gutowski:2004ez,Kunduri:2006ek,Chong:2005hr}. An additional class of non-MG operators should exist to account for the BHs (and perhaps also more exotic entities). We colloquially call all such operators \textit{quantum black holes} since at small dimensions and number of colours we obviously do not expect them to behave as big semi-classical BHs. 
The mathematical problem is that of finding states with protected energy which are not given by products of gravitons. Do they exist? 

A beautiful set of papers \cite{chang2023words,choi2022shape,choi2023towards} recently established their existence. In these works, cohomology classes under the action of a supercharge $Q$ at one-loop were considered, following \cite{XiYinPaper}. By Hodge theorem, each of these classes must contain exactly one BPS operator. Chang and Lin \cite{chang2023words} found the first non-multigraviton cohomology by a careful enumeration at $N=2$. Building on this development, an infinite tower of BH cohomologies at $N=2$ was constructed
in \cite{choi2023towards}. These BHs have no $SU(2)_R$ spin. The classical dimension~$\Delta$,~$SO(6)$~R-charges~$R=R^1= R^2 = R^3 $, and $SU(2)_L$ spin~$J$ of the~$(n+1)^\text{th}$ BH representative $O_n$ are given by
\begin{equation}
    \(\Delta_{(n)},R_{(n)},J_{(n)}\)=\(\tfrac{19}2+4n,\tfrac32,\tfrac52+2n\)
    \label{quantumNumbers} \,.
\end{equation}
We want to go beyond cohomology and really \textit{observe} the quantum BH states. For $n=0$ we managed to find it and cast it in a remarkably compact expression:
\vspace{-0.2cm}
\begin{widetext}
\beqa
|\texttt{BH}\>&=& \left( \psi^1 \phi^{23}\right)  \left( \psi^3 \phi^{23}\right)  \left( \psi^2  \psi^1  \psi^1\right)- \left(\psi^2 \phi^{31}\right)  \left( \psi^3 \phi^{23}\right)  \left(\psi^2  \psi^1  \psi^1\right)+\nn\\
   &+& i Q\cdot\Big[ \tfrac{1}{6}  \left( \psi^1  \psi^2\right)  \left( \psi^2 \psi^3\right)  \left( \psi^3  \psi^1\right)-\tfrac{1}{180}
    \left(f \phi^{23}\right)  \left(f \psi^1\right)  \left(\phi^{31}  \psi^2\right)+\tfrac{1}{20}
    \left(f \phi^{23}\right)  \left(f \psi^2\right)  \left(\phi^{31}  \psi^1\right)+\tfrac{1}{20}
    \left(f \phi^{31}\right)  \left(f \psi^1\right)  \left(\phi^{23}  \psi^2\right)\nn \\
   &-&\tfrac{1}{180}
    \left(f \phi^{31}\right)  \left(f \psi^2\right)  \left(\phi^{23}  \psi^1\right)+\tfrac{2}{45}
    \left(f  \psi^1\right)  \left(\phi^{23} \psi^1\right)  \left(f \phi^{23}\right)+\tfrac{3}{20}
    \left(f  \psi^1\right)  \left(\phi^{23} \psi^2\right)  \left( \psi^1  \psi^3\right)-\tfrac{3}{20}
    \left(f  \psi^1\right)  \left(\phi^{23} \psi^3\right)  \left( \psi^1  \psi^2\right)\nn \\
   &-&\tfrac{1}{9}
    \left(f  \psi^1\right)  \left( \psi^2 \psi^3\right)  \left(\phi^{23}  \psi^1\right)-\tfrac{7}{180}
    \left(f  \psi^2\right)  \left( \psi^1 \phi^{23}\right)  \left( \psi^1  \psi^3\right)+\tfrac{7}{180}
    \left(f  \psi^3\right)  \left( \psi^1 \psi^2\right)  \left( \psi^1 \phi^{23}\right)\Big]+
   \texttt{cyclic}  \, ,\label{N2state}
\eeqa
where $(\,\dots)=\text{tr}(\,\dots)$ and \texttt{cyclic} corresponds to two more terms with R-charge index cyclically shifted by $1$ and $2$. $Q$ is the one-loop action of the supercharge~$Q_{4-}$ which increases the length of the operators by one -- see Appendix \ref{sec:notations}. This state is the precise combination of the cohomology representative of \cite{choi2022shape} and $Q$-exact terms that is killed by the one loop dilatation operator, which was also computed in simultaneous work \cite{Chang:2023zqk}.
\end{widetext}

We suggest that some intuition about these BH states can be gained by embedding them in continuous families of states defined for any $N$, not necessarily integer. After all, the number of colours is effectively related to Netwon's contant so that starting with a bunch of free integrable strings at large $N$ and decreasing $N$ can be though of as increasing the gravitation interaction so that such strings collapse into BHs. We suggest that the MG cohomologies have identically zero anomalous dimensions for all $N$ while the BHs have non-zero energy at large $N$'s which dives down to zero at an integer $N\sim\sqrt{\Delta}$ where they become BPS. Two examples of these two different behaviors are the horizontal lines in figure \ref{figsu2} -- for the MGs -- and the red line in figure \ref{bigFig}  -- for the BH state.

 
\section{Continuous $N$ Hamiltonian and Continuous $N$ Wick Matrix} \la{Ncontinuous}
The one loop dilatation operator acts on gauge invariant operators -- i.e. on polynomials in traces of products of the letters $\mathcal{W}_A$
 (see Appendix \ref{sec:notations}) -- as \cite{beisert, beisertStaudacherKristjansen}
\begin{equation}
    H=\!\!\!\!\sum_{A,B,C,D}\!\!\!\mathcal{C}_{AC}^{BD} (-)^{BC}(-)^{BD}\! :\tr [\mathcal{W}_A,\check{\mathcal{W}}^B\}[\mathcal{W}_C,\check{\mathcal{W}}^D\}:\ .
    \label{hamiltonian}
\end{equation}
The factor $(-)^{AB}$ is $-1$ if both $A$ and $B$ are fermionic and is $+1$ otherwise. The bracket $[X,Y\}=XY-(-)^{XY}YX$ and $\check{\mathcal{W}}^A$ denotes derivative w.r.t the corresponding letter, 
\begin{equation}
    (\check{\mathcal{W}}^A)^a_b (\mathcal{W}_B)^c_d=\delta^A_B \(\delta^a_d\delta^c_b-\tfrac{1}{N}\delta^a_b\delta^c_d\)\ .
    \label{propagator}
\end{equation}
The normal ordering in (\ref{hamiltonian}) stipulates that the derivatives do not act on other fields within the Hamiltonian and, finally, the coefficient $\mathcal{C}_{AC}^{BD}$ is given by the so-called Harmonic action \cite{beisert}. (For an example of the action of $H$ on a simple operator see Appendix~\ref{HexampleAp}.)

The Hamiltonian preserves the total number of letters, but it can split and join traces at subleading order in~$1/N$. By choosing an appropriate basis of multi-traces with some given quantum numbers, we can define a $d\times d$ matrix $\mathbb{H}$ where $d$ is the number of basis elements,\footnote{An example: operators in the $SU(2)$ sector with three scalars~$Z=\phi^{12}$ and three scalars~$X=\phi^{23}$ have a  multi-trace basis with $d=12$ elements:~$\mathbb{v}=\{\tr(X^3Z^3),\tr(X^2ZXZ^2),\!\dots\!,\tr(Z^2)\tr(ZX)\tr(X^2)\}$.
(In this paper \textit{multi-trace basis} = \textit{basis of gauge invariant operators}.) As explained below, for small $N$ this basis is overcomplete.}
\beq
H \mathbb{v}_i=\sum_{j=1}^d \mathbb{H}_{ij} \mathbb{v}_j \,.
\eeq
While (\ref{propagator}) -- and thus (\ref{hamiltonian}) -- only makes sense acting on operators of fixed integer $N$, the whence constructed $d\times d$ matrix $\mathbb{H}$ is a very simple matrix where all entries are simple polynomials in $1/N$ which we can now define for any $N$. We can use it to follow the various physical states -- the $\mathbb{H}$ eigenvectors -- as we adiabatically change the Newton constant $1/N$. 


As we change $N$, the number of independent operators changes due to trace relations. These are identities relating a large single trace operator to products of smaller traces (for instance, in $SU(2)$ we have $\tr X^4\sim\tr X^2 \tr X^2$). Such identities only exist when the number of letters in the operator is bigger than $N$. In order to work with $\mathbb{H}$ uniformly in $N$, we chose a basis at large enough $N$, where trace relations are irrelevant. At low $N$, this basis will be overcomplete but that is not an issue.

Another very useful quantity is what we call the \textit{Wick contraction} matrix $\mathbb{W}$ with entries
\beq
\mathbb{W}_{ij}=\< \mathbb{v}_i(0)^\dagger \mathbb{v}_j(1) \>
\label{innerProduct}
\eeq
computed by simply Wick contracting the fields in $\mathbb{v}_j$ with those in $\mathbb{v}_i$. Once the Wick contractions are done, this matrix is a simple polynomial expression in $N$ which can now be continued to any $N$.

This matrix allows us to define a norm for the states as well as compute some simple extremal correlation functions as explored below.
Note that the norm of an operator that vanishes by trace relations must be zero. At any integer $N$ the problem of finding such operators can be cast as finding the null-space of the matrix~$\mathbb{W}$.

When the theory is unitary, it is possible to pick a basis of operators such that $\mathbb H$ is a Hermitian matrix~\cite{Janik:2002bd,Beisert:2002ff,Casteill:2007td,Gross:2002mh}. More generally, the dilatation operator is self-adjoint with respect to the inner product (\ref{innerProduct})
\beq
\<\mathbb{v}_i^\dagger H\mathbb{v}_j\>=\<\(H\mathbb{v}_i\)^\dagger \mathbb{v}_j\>,
\eeq
or equivalently, $\mathbb{W H}=\mathbb{(W H)^\dagger}$.

The theory at non-integer $N$ does not have to be unitary and we will encounter non-Hermitian $\mathbb H$, complex scaling dimensions and states with negative norm. We will illustrate these phenomena with an explicit example in the next section.

What we have just described for $\mathbb{H}$ and $\mathbb{W}$ is quite general: physical quantities are often computed by diagrams such as Wick contractions or one loop diagrams computing the dilatation operator. Evaluating those diagrams often reduces to counting colour loops and assigning $N$ to each loop. As soon as $N$ is a simple weight factor, we are dealing with so-called Brauer algebras and analytic continuation away from integer $N$ is straightforward -- see~\cite{slava} for a recent exposition\footnote{$U(N)$ is easier to analytically continue compared to~$SU(N)$ since $\varepsilon$ symbols are harder to make sense of for non-integer~$N$ \cite{slava}. We could switch to gauge group $U(N)$ if faced with related obstacles.} and Appendix~\ref{su2ap} for a simple example.\footnote{Further examples of similar analytic continuations include continuing $O(n)$ vector models to $n\to 0$ to study self-avoiding walks \cite{de1972exponents}, $\mathbb{Z}_n$ Potts models to $n\to 1$ to study percolation \cite{fortuin1972random}, $\mathbb{Z}_n$ flavored QFT's to $n\to 1$ to define entanglement entropy from Reyni entropies, $U(N)$ matrix models with $N\to 0$ to determine a class of large operators in $\mathcal{N}=4$ SYM \cite{Bargheer:2019kxb} and even $N\to - N$ in AdS/CFT examples to try to unveil dS/CFT \cite{anninos2016higher} etcetera. 
}

We conclude this section with an observation about MG operators. The MG representatives found in \cite{XiYinPaper} are $N$-independent and correspond to distinct BPS states as long as trace relations can be ignored. When trace relations kick in, the cohomologies can become $Q$-exact, which in the state picture corresponds to BPS states being killed by trace relations. So, for all integers $N$ bigger than a threshold, the MG states have protected dimensions. Given that the 1-loop anomalous dimensions are eigenvalues of a finite dimensional matrix $\mathbb{H}$ whose entries are simple polynomials in $1/N$, they must be identically zero for any $N$.

\section{Two Scalars Toy Model}

To illustrate the effects alluded to in the previous section, let us look at a simple example in the so-called $SU(2)$ sector, with $Z \equiv \phi^{12}$ and $X \equiv \phi^{23}$. 

Consider operators made out of three $X$'s and three $Z$'s such as $\{\tr Z^3X^3, \tr Z^2X\tr ZX^2,\ldots\}$. There are 12 such operators. The Hamiltonian has additional symmetries we can leverage: it commutes with the $SU(2)$ raising operator~$J_+$, which flips an $X$ to $Z$, and with parity, 
\beqa
\Pi: \tr\mathcal{W}_1\ldots\mathcal{W}_L\mapsto(-1)^{L+n_f(n_f-1)/2}\tr\mathcal{W}_L\ldots\mathcal{W}_1\ ,
\label{parity}
\eeqa
where $n_f$ denotes the number of fermions in the the trace.
The $12\times 12$ matrix $\mathbb{H}$ can be further block decomposed according to these symmetries. 

\begin{figure}[t]
    \centering
    \includegraphics[scale=0.3]{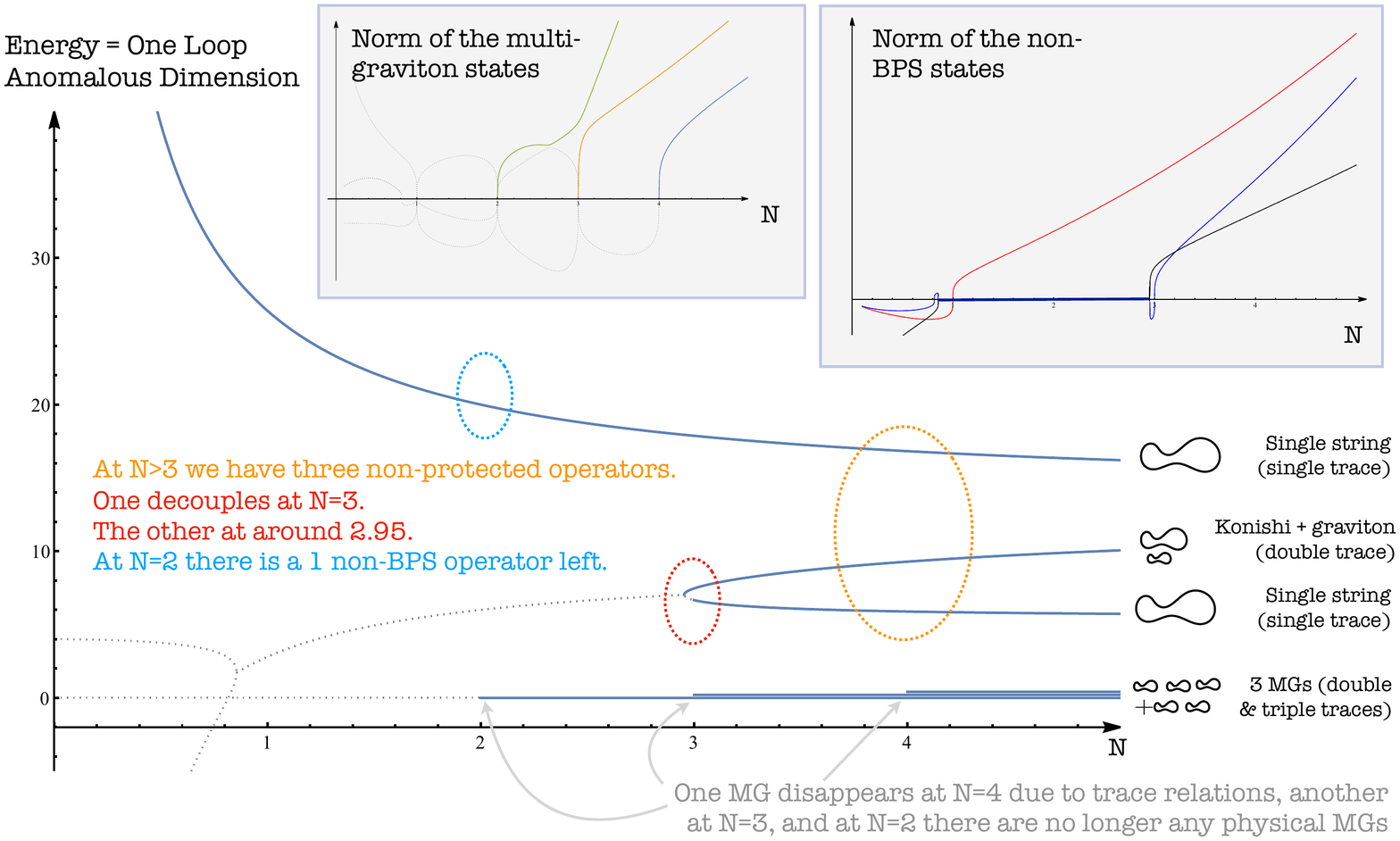}
    \vspace{-1.5cm}
    \caption{At large $N$ there are 3 MGs and 3 non-protected states. This counting persists for all~$N\ge 6$. As we decrease~$N$ some of these states start decoupling (drawn as dashed lines in the figure) due to trace relations. At $N=2,3,4$ we have $1,3,5$ physical states respectively -- with a positive norm and non-negative energy.}
    \label{figsu2}
\end{figure}

One nice block is given by the set of operators with even parity and killed by $J_+^2$ -- we call them next-to-highest weight (NHW) since highest weight states are those killed by a single $J_+$. This forms a 6 dim subspace, see Appendix \ref{su2ap}. The mixing matrix in the NHW basis~
\begin{equation}
\mathbb{H}=\left(
\begin{array}{cccccc}
 4 & -4 & \frac{40}{N} & \tfrac{16}{N} & -\tfrac{8}{N} & 0 \\
 4 & 16 & -\tfrac{20}{N} & \tfrac{16}{N} & \tfrac{12}{N} & 0 \\
 0 & 0 & 0 & 0 & 0 & 0 \\
 0 & 0 & 0 & 0 & 0 & 0 \\
 \tfrac{16}{N} & \tfrac{24}{N} & 0 & 16 & 12 & \tfrac{40}{N} \\
 0 & 0 & 0 & 0 & 0 & 0 \\
\end{array}
\right) \, .
\end{equation}
Such a simple matrix can even be diagonalized by hand. It has three identically zero eigenvalues -- which correspond to MG states -- and three non-zero energies $\gamma(N)$ given by the roots of a simple cubic equation 
\beq
\gamma ^3-8 \gamma ^2 N+\gamma  \left(20 N^2-10\right)+5 N \left(2-3 N^2\right)=0 \,, \label{eq:gamma}
\eeq
defining a Riemman surface where the energies live. In figure \ref{figsu2} we plot these eigenvalues as a function of $N$. We nicely see two trajectories meeting at a square root branch point near $N=3$ and becoming complex for smaller values of $N$ (see the red circle).\footnote{These points where the energies collide and become complex are very close to integers. The one in the figure happens at $N\simeq 2.95$ and in the BH plot of figure \ref{bigFig} we find even more striking numerical coincidences with plenty of colliding points immediately to the left of integers where trace relations kick in.}

As described in the Appendix \ref{su2ap}, the three massive states at large $N$ split up as (i) a double trace made of an $SU(2)$ representative of Konishi dressed by a graviton
\beqa
|\psi\>^{N=\infty}_{\gamma=12}=|\texttt{graviton}\rangle \times |\texttt{su(2) konishi}\rangle \label{konishiSU2} \, ,
\eeqa
and (ii) two single traces. At large $N$ strings become free and states with different numbers of traces do not mix. Once we analytically continue in $N$ they become connected, different sheets of a same unifying surface (\ref{eq:gamma}).

In figure \ref{figsu2} we find complex and negative energies. Aren't these the eigenvalues of a positive definite Hermitian operator $H=\{Q,Q^\dagger \}$ which should thus only have positive eigenvalues? What is going on? 

The solution to this small puzzle is that trace relations are kicking in and saving us from non-unitarity contradictions. As discussed above, a convenient object to automatically encode these relations is the Wick contraction matrix $\mathbb{W}$ which in this case is a simple~$6\times 6$ matrix 
\beq
\!\!\!\mathbb{W}\!=\!\! \footnotesize\left(\!
\!\begin{array}{ccc}
 4 N^6\!-\!4 N^4\!+\!36 N^2\!-\!36 & 6 N^6\!+\!14 N^4\!+\!4 N^2\!-\!24 &\! \dots \\
 6 N^6\!+\!14 N^4\!+\!4 N^2\!\!-\!24 & 14 N^6\!+\!26 N^4\!-\!24 N^2\!\!-\!16 & \!\dots \\ \vdots & \vdots & \!\ddots 
\end{array}
\!\!\right)\! . \la{Wsu2}
\eeq
This is a symmetric real matrix so that the norm of any state $\mathbb{v}^\dagger \cdot \mathbb{W} \cdot \mathbb{v}$ is real and the eigenvalues of $\mathbb{W}$ are also real. At large $N$ they are all positive but as we decrease $N$ we see that the norm of the non-protected states can become negative or even vanish in a finite $N$ segment. This second scenario can happen for states whose energy acquires an imaginary part as illustrated in the right-most inset in figure \ref{figsu2}. In the second inset of that figure we plot the three eigenvalues of the $3\times 3$ restricted Wick matrix once we project it onto the MG degenerate subspace. Precisely at integers where trace relations kick in, some eigenvalues of this matrix disappear and the corresponding states become unphysical.

In sum: if we stick to integer $N$'s and 
systematically throw away unphysical states which drop out due to trace relations then all energies and norms will be nice non-negative real numbers. If, on the other hand, we keep all states and allow $N$ to take continuous values we gain a more unified description of the spectrum where many previously unrelated states become nicely connected.

\section{The BH sector} \la{sec:BH}

\begin{figure*}[t]
  \includegraphics[width=0.95\textwidth]{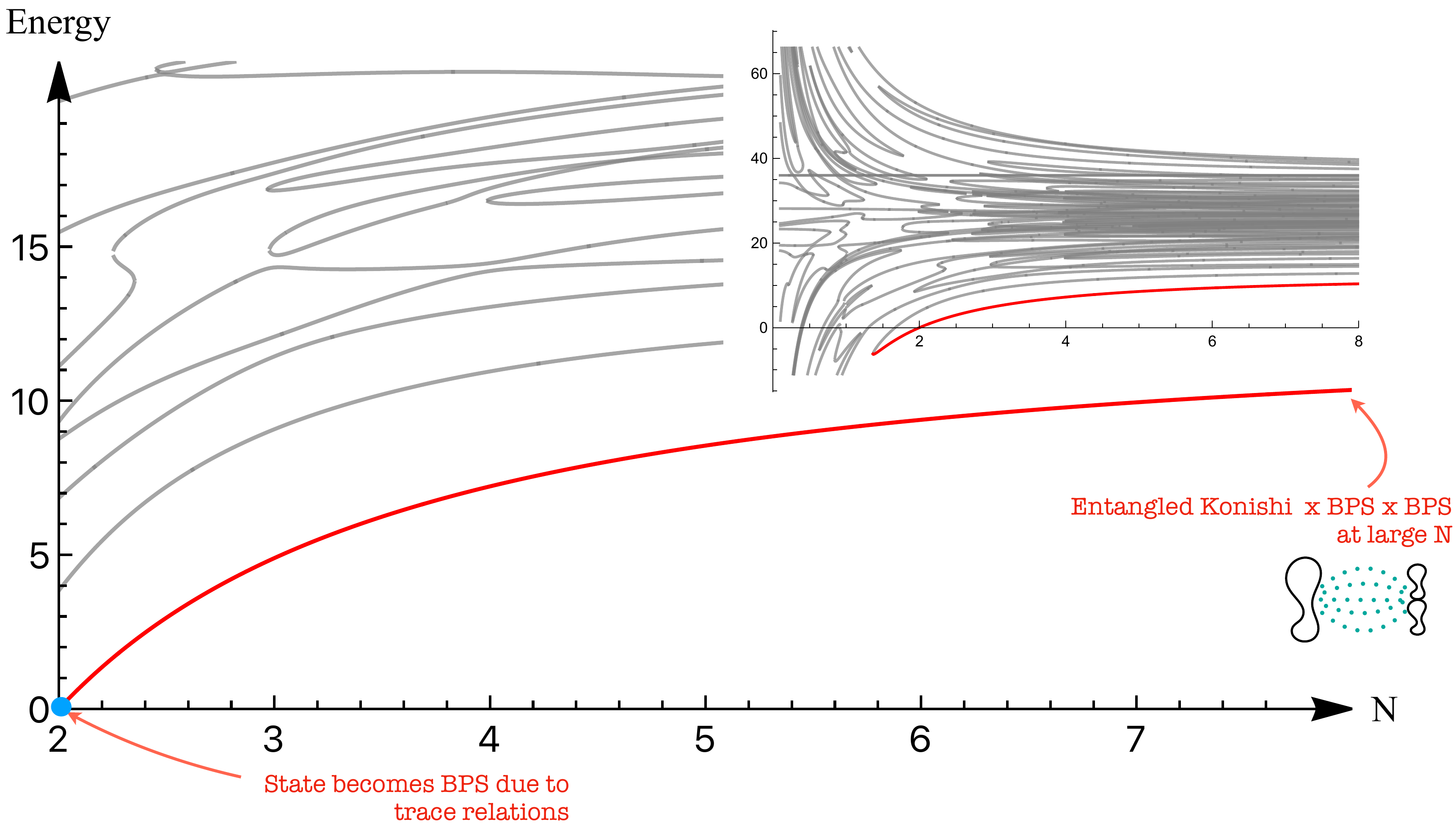}
  \caption{Energies of the 69 states in the \textit{BH sector} described in the main text. 
  There are several level repulsions (and some accidental crossings which will likely get resolved at two-loops, see Appendix \ref{crossingSU2}). Some things are similar to the two scalars toy example -- for instance, states annihilating  extremely close to integers where the trace relations kick in -- but there is an obvious novelty here: There is a state whose energy goes to zero at $N=2$. This is the so-called $1/16$-th $N=2$ black hole state identified by the blue circle in the figure. This state flows to an entangled triple trace of a Konishi multiplet operator and two gravitons at large~$N$~(\ref{konishiBH}). The inset shows the same plot in a bigger range and interestingly the black hole annihilates with the second excited state -- which is a single trace at large $N$ -- a bit to the left of $N=1$.} \label{bigFig}
\end{figure*}

Multiplets that correspond to BPS BHs cannot have identically vanishing energies for all $N$ -- this is what distinguishes them from MGs. Together with analyticity of one-loop energies, this implies that a BH multiplet in the $SU(N_0)$ theory must develop an anomalous dimension when~$N$ is perturbed away from~$N_0$. However, as we change the coupling while staying at~$N=N_0$, the multiplet is expected to remain short\footnote{In general $\mathcal{N}\!=\!1$ theories, loop corrections to $Q$ can modify the 1/4-BPS spectrum \cite{Budzik:2023xbr}. We verified by explicit computation that the first higher loop correction of \cite{Budzik:2023xbr} preserves the BH cohomology~$O_0$.} \cite{Grant:2008sk,chang2023words}. 

\setlength{\skip\footins}{26pt}

One possible scenario is as follows. Assuming that the representation continues to be unitary away from~$N_0$, the BH multiplet recombines with another short multiplet (which is killed by $SU(N_0)$ trace relations). Such BPS BH multiplets lie at the unitarity threshold and satisfy shortening conditions $A_1$ or $A_2$ \cite{cordova2019multiplets, Dolan:2002zh} (see Appendix \ref{sec:rec}).
The simplest BH multiplet discussed below, and all others found in \cite{choi2023towards} are of the type $A_1\bar{L}$ and seem to realize this scenario. It would be interesting to investigate if there are examples of BH multiplets which become non-unitary away from $N_0$.

Recently in \cite{chang2023words}, the first non-MG cohomology was found in the $SU(2)$ theory. Its quantum numbers are given by (\ref{quantumNumbers}) with $n=0$. {With these quantum numbers, at tree level, the operators are made out of letters without $SU(2)_R$ dotted indices, see Appendix \ref{sec:notations}. 
At large $N$, there are a total of 8013 multi-trace operators with these quantum numbers. We can reduce the number of states we are dealing with, by imposing certain conditions: The BH is a primary under $\textrm{psu}(1,2|3)$ and it is a singlet under the unbroken $SU(3)$ of R-symmetry $SO(6)\sim SU(4)$ (if this were not the case, we should have seen additional non-MG states at this level in the exhaustive search of~\cite{chang2023words}). We thus impose
\begin{align}
   S^i_+ \ket{\psi }&=0\ , \,i=1\ldots 3\nn\\ 
    R_i^j\ket{\psi}&=0\ , \, i\neq j,\ i,j=1,\ldots,3 \, , 
\end{align}
where in the first line, we only require that the classical $S$'s kill the states.\footnote{{$\tilde S_{i\dot\alpha}\ket{\psi}=0$ is automatically satisfied in the no-derivative sector.}} These constraints commute with the action of $H$. We can further separate the states in this subspace by their parity (\ref{parity}) -- there are 69 parity even states and 30 parity odd states. Diagonalizing the mixing matrix $\mathbb{H}$ in the parity even sector, gives us the anomalous dimensions in figure \ref{bigFig}. 
The lowest energy trajectory in this sector corresponds to the BH. As anticipated above, the state becomes $\tfrac1{16}$-BPS at $N=2$ because a part of the multiplet becomes null due to $N=2$ trace relations. At $N=2$ it is given by (\ref{N2state}).\footnote{More precisely, at $N=2$ it \textit{can} be written as in (\ref{N2state}). Due to trace relations there are many equivalent representations. A canonical choice comes from diagonalizing $\mathbb{H}$ at general $N$ and analytically continuing the eigenvector to $N=2$ as in figure~\ref{bigFig}. This yields a way more involved representation, with $\sim 8K$ terms, see Appendix~\ref{entanglementAp}.}  
At large $N$, it takes the form of an interesting entangled triple trace,\footnote{At large $N$ we can find not only this state but \textit{all} 69 states. They break into $25$ single traces, $36$ double traces and $8$ triple traces at infinite~$N$. We can compute the energy of all of them using integrability. The hardest are the single traces since they contain the longer traces but even those can be nicely computed using the QQ-system technology developed in \cite{dimapapers1,dimapapers2,dimapapers3}. With Dima Volin's help, we found all infinite $N$ energies and obtained a perfect agreement with the large $N$ asymptotes of figure~\ref{bigFig}.}
\beq
\underbrace{\(\tr fX+\tr\psi^3\psi^2\)^2}_{\text{BPS}\times \text{BPS}}\overleftrightarrow{\mathbf{E}}_1\underbrace{(\tr ZY\psi^1-\tr YZ\psi^1)}_{\text{Konishi}} + \texttt{cyclic}\ .
\label{konishiBH}
\eeq
On the left, we have (a product of) two gravitons and on the right we have a Konishi (multiplet operator). The entangling operator $\overleftrightarrow{\mathbf{E}}_1$ adds some charges on the right and removes the same ones from the left. For example, it will contain a term $\tfrac{1}{6}  \overleftarrow{S^1}  \otimes \overrightarrow{Q_1}$ 
where $Q_1$ acts on the right and its conjugate $S^1$ acts on the left. For compactness we will write it as $\frac{1}{6}Q_1$ leaving implicit the conjugate action on the left. The full entanglement operator is then 
\begin{widetext}
\beqa
\overleftrightarrow{\mathbf{E}}_1=&&1+\tfrac{1}{6}Q_1+\tfrac{1}{2}Q_2+\tfrac{1}{2}Q_3-\tfrac{1}{12}Q_2 Q_1-\tfrac{1}{12}Q_3 Q_1-\tfrac{1}{6}Q_3 Q_2-\tfrac{1}{2}R_1^2-\tfrac{1}{4} R_1^2 Q_3+\tfrac{1}{36} R_1^2 Q_3 Q_2+\tfrac{1}{24}
   R_1^3 Q_1 Q_3+\nn\\
   &&\tfrac{1}{36} R_1^3 Q_2 Q_3+\tfrac{1}{24} R_2^3 R_1^2 Q_3+\tfrac{1}{24}
   R_3^2 R_1^3 Q_2+\tfrac{1}{24} Q_2 Q_3 R_1^3+\tfrac{1}{24}
   Q_3 Q_2 R_1^2-\tfrac{1}{216} R_1^3 R_1^2 Q_3 Q_2-\nn\\
   &&\tfrac{1}{48} R_3^1 Q_3 R_1^3 R_1^3+\tfrac{1}{72}
   Q_3 Q_2 R_1^2 Q_1-\tfrac{1}{36}
   Q_3 Q_2 Q_1-\tfrac{1}{48}
   R_2^1 Q_2 R_1^2 R_1^2+\tfrac{1}{24}
   R_3^1 R_2^1 R_1^3 R_1^2 Q_1\ , \la{Eop}
\eeqa
where $Q_i\equiv Q_{i-}$,  $\overleftrightarrow{\mathbf{E}}_2$ and $\overleftrightarrow{\mathbf{E}}_3$ are defined by cyclically changing $1\rightarrow2\rightarrow3\rightarrow1$, and each term 
is implicitly multiplied by its conjugate (obtained by $Q_i\rightarrow S^i$ and $R^a_b\rightarrow R^b_a$ and flipping the order of operators) acting on the left.
\end{widetext}

~\newline
\vspace{-22pt}

Perhaps the most famous quantum state, the EPR state, can be nicely entangled in a similar fashion,  
\beq
\ket{\uparrow}\overleftrightarrow{\mathbf{e}}\ket{\downarrow}=\ket{\uparrow\downarrow}+\ket{\downarrow\uparrow} \,, \qquad \overrightarrow{\mathbf{e}}=1+\overleftarrow{\sigma_-} \otimes \overrightarrow{\sigma_+} \,,
\eeq
which is what motivated us to dub $\overleftrightarrow{\mathbf{E}}$ as the \textit{entanglement} operator. Amusingly, in the previous section we encountered Konishi dressed by gravitons as well but a key difference is that there is no entanglement in (\ref{konishiSU2}) while there is a huge amount in (\ref{konishiBH}). 

A cute -- albeit useless -- graphical representation of the BH state at $N=2$ and $N=\infty$ is given in Appendix~\ref{entanglementAp}.


\section{The $1/16$-th BH Fate} \la{sec:bhfate}

We could now try to follow the large $N$ state (\ref{konishiBH}) to large $\lambda$. Here we have to be careful. 

The easy thing to do is to follow the state to strong coupling at $N=\infty$. This corresponds to free strings whose energies are additive so that the energy of the state~(\ref{konishiBH}) is equal to the energy of Konishi plus the energy of the two gravitons. It is thus given by $19/2+\gamma_\texttt{Konishi}(\lambda)$ because only Konishi develops an anomalous dimension. 
Using integrability, $\gamma_\texttt{Konishi}(\lambda)$ can be followed all the way from weak to strong coupling \cite{Gromov:2009zb}. At weak coupling it is given by $12\lambda+\dots$ while at strong coupling it behaves as $2\lambda^{1/4}-4+2/\lambda^{1/4}+\dots$ \cite{GKP,AFS,Gromov:2009zb}. In sum, if we follow our black hole state (\ref{konishiBH}) to strong coupling at infinite $N$ we end up at a heavy state with energy 
\beq
E_\texttt{Integrability}=2 \lambda^{1/4}+11/2+\dots \label{Eintegrability}
\eeq
as depicted by the solid red line in figure \ref{bhfate}. 


\begin{figure}[t]
    \centering
    \includegraphics[width=0.5\textwidth]{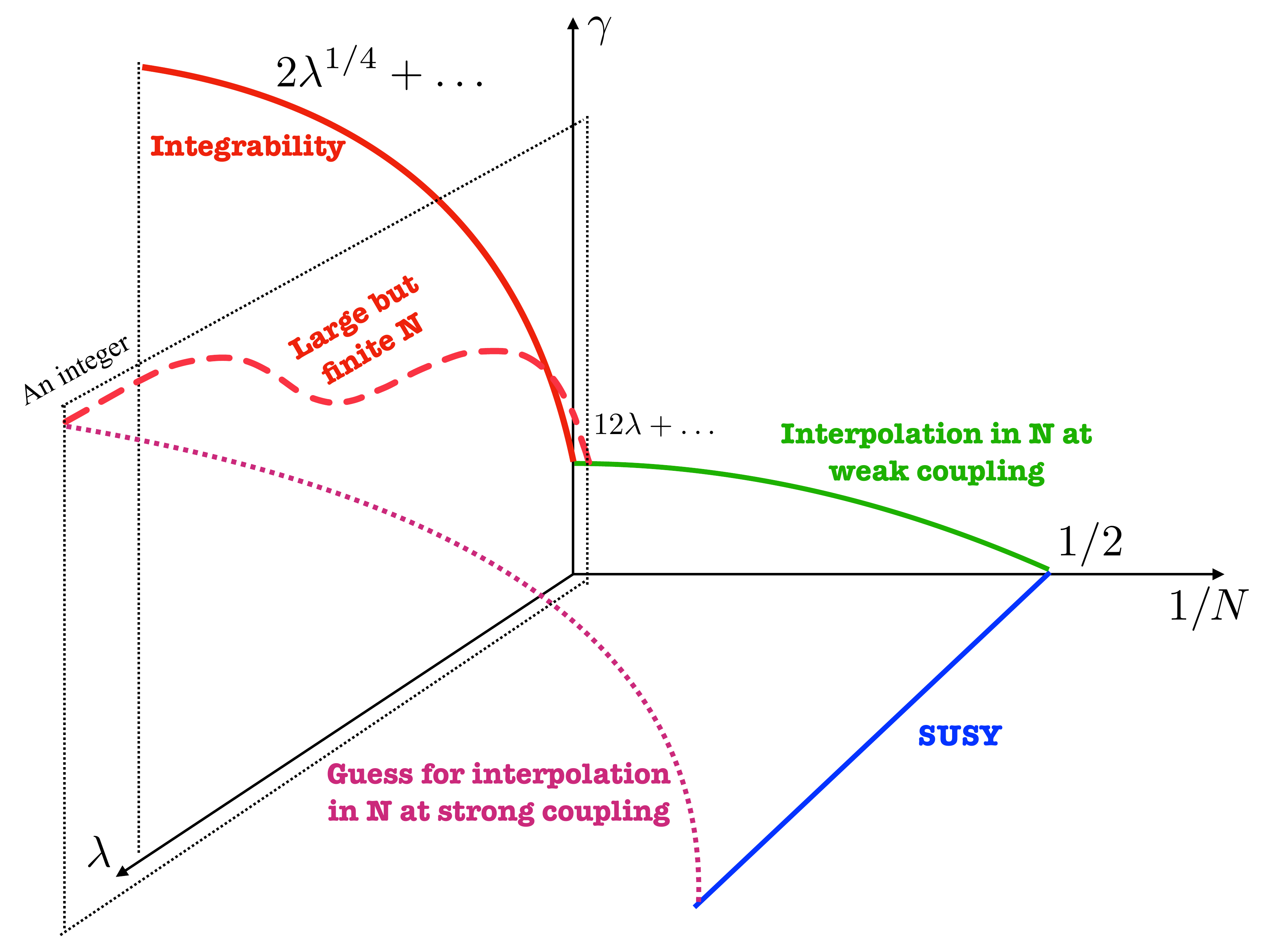}
        \caption{At $N=2$ the BH state is a $1/16$-BPS state whose energy is zero at any coupling (blue line). The green line is a sketch of the weak coupling interpolation in section~\ref{sec:BH}. At infinite $N$ the BH state is a triple trace state made of Konishi and two gravitons. As we crank up the coupling at infinite~$N$, the anomalous dimension follows Konishi \cite{Gromov:2009zb} (red solid line). If we stick to large but \textit{finite} $N$ as we go to strong coupling, the state will probably follow a stair-case pattern like figure \ref{stairs} and eventually plateau at an approximate MG state (dashed red line). The energy at strong coupling will be a finite integer away from the classical dimension $19/2$ of the BH state. Then, if we follow the state to small $N$ it should again dive to zero at $N=2$ as depicted by the dotted pink line; it would be important to find a new idea which would allow us to say something quantitative about this strong coupling interpolation. }
    \label{bhfate}
\end{figure}

The harder -- but somewhat more accurate -- thing to do is to follow the state at large but finite $N$.  The final result can be dramatically different since at finite $N$ strings will no longer be free, integrability is broken, and level repulsion will dramatically affect the fate of the various states. The point is that there are two families of states which will clearly cross at infinite $N$:
\begin{itemize}
    \item Products of protected operators, i.e. operators like $\Tr(ZZ)\Tr(\bar Z\bar Z)$. At infinite $N$, and for any coupling, these operators will have no anomalous dimension. At large but finite $N$ they acquire a small binding energy $\gamma(\lambda)$ of order $1/N^2$. They are schematically depicted as the red dotted lines in figure \ref{stairs}. They are MG-like states.\footnote{These states with a small binding energy differ from genuine MGs  discussed in the previous sections such as $\Tr(ZZ) \Tr(Z Z)$ which have identically zero energy independently of $N$ and $\lambda$.} 
    \item Products involving at least one non-protected state such as Konishi -- such as our state (\ref{konishiBH}). As soon as a state involves at least one non-protected state (dual to an excited string) its energy will grow as $\lambda^{1/4}$ at infinite $N$ as described in the previous paragraph, see (\ref{Eintegrability}). These states are represented by the blue solid lines in figure \ref{stairs}. They necessarily cross many red dotted lines at finite values of $\lambda$. 
\end{itemize}
At large but finite $N$ the level crossing will be resolved by nonperturbative mixing\footnote{In perturbation theory (PT), loop corrections to the dilatation operator commute with the classical dilatation \cite{beisert}. For a more conventional type of level crossing, which \textit{is}  resolved in PT, see App.~\ref{crossingSU2}.} \cite{Korchemsky:2015cyx} as sketched on the right panel of figure \ref{stairs}. A small excited string at weak coupling like Konishi -- or Konishi dressed by a bunch of other strings as in our case (\ref{konishiBH}) -- will therefore \textit{not} grow into a heavy massive excited string with huge energy at strong coupling but instead end up as a MG-like state after some level repulsion(s).\footnote{For example, the lightest non-protected operator, the Konishi singlet $O_\texttt{Singlet Konishi}=\Tr(Z\bar Z)+\Tr(X\bar X)+ \Tr(Y\bar Y)$ with weak coupling dimension $2$ will mix with a MG-like state of the schematic form $\Tr(ZZ)\Tr(\bar Z\bar Z)+ \dots$ at some finite coupling $\lambda$ and follow this state from this point on. Its \textit{true} strong coupling dimension will thus be close to $4$. Without level crossing, Konishi must remain the lightest state at any value of the coupling.  } In sum, at large but finite $N$ we expect our state to end up as a MG-like state with an energy some finite integer distance away from its weak coupling energy $19/2$, 
\beq
E_{\text{\st{\texttt{Integrability}}}}=19/2+\texttt{integer} \label{EnoIntegrability}\,
\eeq
as sketched by the dashed red line in figure \ref{bhfate}.\footnote{The staircase part of this cartoon can be worked out semi-rigorously. We simply need to figure out how many multi-trace operators are there with the BH quantum numbers and how many non-protected states with energies smaller or equal to the BH exist. Then it is a simple matter of plotting the energies of all these states using integrability and then resolving the several level crossings (by hand). This would tell us precisely to what half-integer energy will the BH go to at strong coupling. 
}

\begin{figure}[t]
    \centering
    \includegraphics[scale=0.3]{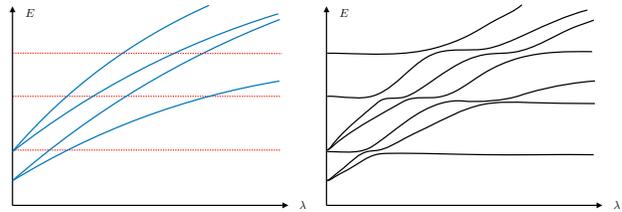}
    \vspace{-3cm}
        \caption{On the left cartoon, the blue solid lines represent the planar dimensions of non-protected states -- whose energy blows up at strong coupling -- and the red dashed lines correspond to products of protected traces (gravitons). At large but finite $N$ mixing will resolve the level crossing and we end up with something as  sketched on the right. (Not all horizontal lines get repelled - some stay protected by SUSY and those are precisely the MGs; we are not drawing those.) 
        }
    \label{stairs}
\end{figure}


One could be puzzled: How can it make such a difference to follow the state to strong coupling at huge $N$ versus infinite $N$? It depends what we mean by \textit{follow}. If we imagine a real experimental setup where we slowly tune the coupling in time, interpolating between  weak and strong coupling, then we would stay in the lowest energy state -- the dashed line in figure \ref{bhfate} -- if we move slowly enough. If we move too fast, we will jump levels and follow the integrability prediction -- the solid red line.  
The larger $N$, the slower we would need to move nearby a level repulsion to stay in the lowest level, see Appendix \ref{adiabatic} for an illustrative toy model.



\section{Discussion}
Once we convert gauge theory computations into loop diagrammatics where the number of colours shows up as a simple weight factor, we can take it to be $N=3.14159$ or any other value. This can be useful if we want to follow adiabatiacally operators across $N$ from regimes where we understand them very well (the integrable $N\to \infty$ corner for example) all the way to more challenging domains. 

We used this idea to provide a new perspective on the lightest of the recently discovered non-multigraviton $1/16$-BPS cohomologies, sometimes simply called the (lightest) \textit{black hole} state.\footnote{We discussed the fate of this state at week and strong coupling in sections \ref{sec:BH} and \ref{sec:bhfate} respectively.} At weak coupling, it is a protected state with zero anomalous dimension at $N=2$ while at large $N$ it is a non-trivial, very entangled, triple trace state dual to a Konishi string plus two gravitons. 

Was this expected? Is it more natural that the BH comes from such a ``maximally fragmented" triple trace product of light strings instead of a single large string? Should we expect something similar for the other recently found BHs~\cite{choi2023towards}? Will the $n=1$ state in (\ref{quantumNumbers}) flow to a quadruple trace with one small non-protected string dressed by three gravitons?\footnote{We did manage to find the analogue of (\ref{N2state}) for $n=1$ by brute force study of the $\mathbb{H}$ null space but we did not manage to follow it in~$N$ yet as in figure \ref{bigFig}; it is computationally expensive.} 

What about $N>2$ black holes? It would be interesting to find a BH state for~$N=3$ and follow it again in~$N$. An interesting question is whether it will remain a 1/16-BPS BH at $N=2$. Naively this seems plausible: decreasing $N$ further from $3$ to $2$ corresponds to increasing the gravitational force further so why would the black hole go away?\footnote{Well, it could remain a black hole but not a supersymmetric one...} If that is the case, its energy should decrease to zero at $N=3$ from the right but then, instead of going negative -- and decoupling from the theory -- it should have have a double zero there, jump back up to dive back to zero one unit to the left again, at $N=2$.\footnote{A more esoteric scenario would be to have states becoming unphysical and then physical again. For instance, a state could have a negative norm between~$2$ and~$3$ but be physical at both~$2$ and~$3$. We never saw anything like this in our explorations. Would be interesting to either find one such example or rule this scenario out from first principles.} It would be very cool to see this happen! It would also be very interesting to continue the superconformal index in $N$ (possibly using recent advances in the giant graviton expansion \cite{Gaiotto:2021xce,Imamura:2021ytr,Choi:2022ovw}) -- what happens to the BH contribution at non-integer $N$?

We also discussed the fate of the BH state as we move both to large $N$ and strong coupling. We argued that if we move slow enough we end up with a gas of gravitons at strong coupling (four gravitons for the lightest BH studied in this note) but if we move quickly we will end up with a very massive small string (the Konishi string) dressed by two gravitons, see figure \ref{bhfate}. As depicted there, there should also be an interpolation line across many~$N$'s at strong coupling. It would be fascinating to try to work it out. Perhaps some localization examples could be a good starting point to develop some intuition.

Finally, we should probably not be obsessed with the BPS states as far as black hole physics goes. What we are calling \textit{the BH state}, the lowest curve in \ref{bigFig}, is a priori not more of a black hole than any of the other energy levels. In Appendix \ref{entanglementAp} we started a simple exploration of these other states and their statistics. We find that the correlation functions with simple probes of these states lie close to typicality, as expected for heavy states \cite{Balasubramanian:2005mg, Balasubramanian:2007qv}. It would be very interesting if we could make contact between the finite $N$ partition function of \cite{Aharony:2003sx,Kristensson:2020nly} and some of the physics of non-supersymmetric BHs such as those found by Chong, Cvetic, Lu and Pope \cite{Chong:2005hr}, from a microscopic description of these objects as quantum operators in maximally supersymmetric gauge theory.

\begin{acknowledgments}

We are specially grateful to Davide Gaiotto for numerous enlightening discussions and valuable comments on the draft. We thank Jacob Abajian, Nathan Berkovits, Frank Coronado, Alexandre Homrich, Ji Hoon Lee, Juan Maldacena, Enrico Olivucci and Dima Volin  for useful discussions and Chi-Ming Chang, Li Feng, Ying-Hsuan Lin and Yi-Xiao Tao for coordinating the submission to the arXiv of related work \cite{Chang:2023zqk}. 
Research at the Perimeter Institute is supported in part by the Government of Canada 
through NSERC and by the Province of Ontario through MRI. This work was additionally 
supported by a grant from the Simons Foundation (Simons Collaboration on the Nonperturbative Bootstrap \#488661) and ICTP-SAIFR FAPESP grant 2016/01343-7 and FAPESP grant 2017/03303-1. Part of the research was completed at the KITP workshop “Bootstrapping Quantum Gravity". This work was supported in part by the National Science Foundation under Grants No. NSF PHY-1748958 and PHY-2309135.
\end{acknowledgments}

\appendix 
\section{Notation and Conventions}
\label{sec:notations}
The field content of $\mathcal{N}=4$ SYM consists of the following fields in the adjoint of $SU(N)$: six real scalars $\phi^{ij}$, four chiral and four anti-chiral fermions $\Psi^i_{\alpha}$ and $\bar\Psi_{i \dot\alpha}$ and the gauge bosons $\mathcal{A}_\mu$. The R-charge indices take the values $i,j=1\ldots4$, and the $SU(2)_L$ and $SU(2)_R$ Lorentz indices $\alpha, \dot\alpha=\pm$. We can define the self-dual and anti self-dual field strengths $\mathcal{F}_{\alpha\beta}$ and $\mathcal{\tilde{F}}_{\dot\alpha \dot\beta}$. 

In this paper, we consider 1/16-BPS operators. Let $Q\equiv Q_{4-}$ and $S\equiv S^{4-}$ be the charges that kill them. Such operators must be made out of the following BPS letters, each of which is killed by the classical action of $Q$ and $S$, 
\begin{itemize}
\setlength\itemsep{3pt}
    \item $(X,Y,Z)\equiv\phi_{4i}=\tfrac12\epsilon_{4ijk}\phi^{jk}$,
    \item $\psi^i=\Psi^i_+$ and $\lambda_{\dot\alpha}=\bar\Psi_{4\dot\alpha}$,
    \item $f=\mathcal{F}_{++}$,
    \item Covariant derivatives $D_{+\dot\alpha}$\ .
\end{itemize}
At one-loop, $Q$ acts on the BPS letters as follows,
\beqa
\{Q,\psi^m\}&=&-i \epsilon^{mnp}[\phi_{4n},\phi_{4p}]\nn\\
\[Q,f\] &=& i[\phi_{4n},\psi^n]\nn\\
\[Q,D_{+\dot\alpha}\] &=& -i[\lambda_{\dot\alpha},\ \} \, .
\eeqa
There is a smaller subspace in which we can look for 1/16-BPS operators -- we restrict to operators which have no $SU(2)_R$ indices (i.e. no $\dot\alpha$'s). It is built out of the following seven letters
\begin{equation}
    \mathcal{W}_A=\{X,Y,Z,\psi^1,\psi^2,\psi^3,f\} \, .
    \label{lettersNoDer}
\end{equation}
This subspace is closed at one-loop because the Beisert Hamiltonian only shuffles the indices around and does not add new ones. We call this the ``No derivative sector".\footnote{This is the same as the ``BMN sector" discussed in \cite{choi2023towards}.} The black hole discussed in the main text lies within it.

\section{Non-planar H -- Indices and Signs}
\label{HexampleAp}
In \cite{beisert}, the one-loop Hamiltonian was written down. However, there are some sign ambiguities which were not completely spelled out -- see footnote 1, Chapter 3 of \cite{beisert}. Since we are forced to include all the letters in (\ref{lettersNoDer}) to find 1/16-BPS states, we need to fix all the signs.

The scaling dimensions of the generators of the psu$(2,2|4)$ do not receive quantum corrections. So, for algebra generators $\mathfrak{g}$, we have
\beqa
\[\delta D(g_{\text{YM}}), \mathfrak{g}(g_{\text{YM}})\] & = & 0\ ,\nn\\
\implies g_{\text{YM}}^2 \[ D^{(2)},\mathfrak{g}^{(0)}\] &=& 0\ .
\eeqa
In other words, $H$ commutes with the classical generators. By imposing this condition, we fixed the sign ambiguities and ended up with the following expression for~$H$,
\begin{equation}
    H=\sum_{A,B,C,D}\mathcal{C}_{AC}^{BD} (-)^{BC}(-)^{BD} :\tr [\mathcal{W}_A,\check{\mathcal{W}}^B\}[\mathcal{W}_C,\check{\mathcal{W}}^D\}:\nn 
\end{equation}
Let us now work out a simple example that shows how to work with the dilatation operator. Consider the rank one sector $SU(1|1)$ that consists of one boson, say $Z\equiv\phi^{12}$, and one fermion, say $\psi\equiv \Psi^1_+$. In this sector, the coefficient $\mathcal{C}_{AC}^{BD}$ in (\ref{hamiltonian}) is simply a super-permutation, i.e.,
\beqa
\mathcal{C}_{Z\psi}^{Z\psi}=\mathcal{C}_{\psi Z}^{\psi Z}=
-\mathcal{C}_{Z\psi}^{\psi Z}=-\mathcal{C}_{\psi Z}^{Z\psi}=\tfrac{1}{2}\mathcal{C}_{\psi\psi}^{\psi\psi}=1
\label{su(11)} \, .
\eeqa
Plugging these into (\ref{hamiltonian}), we get the following form for $H$ in this subsector
\beq
H = \frac{-2}N \(\tr [Z,\psi][\check{Z},\check{\psi}] -2 \tr\(\psi\psi\check{\psi}\check{\psi}\) \) \, .
\label{Hsu11}
\eeq
Consider the action of $H$ on the state $\tr (\psi\psi Z)$. For ease of illustration, let us consider the dilatation operator in the $U(N)$ theory. The propagator in this case is just the first term in (\ref{propagator}). The first term in (\ref{Hsu11}), $\tr (Z\psi\check{Z}\check{\psi})$, acts on the state as,
\beq
\includegraphics[width=0.5\textwidth]{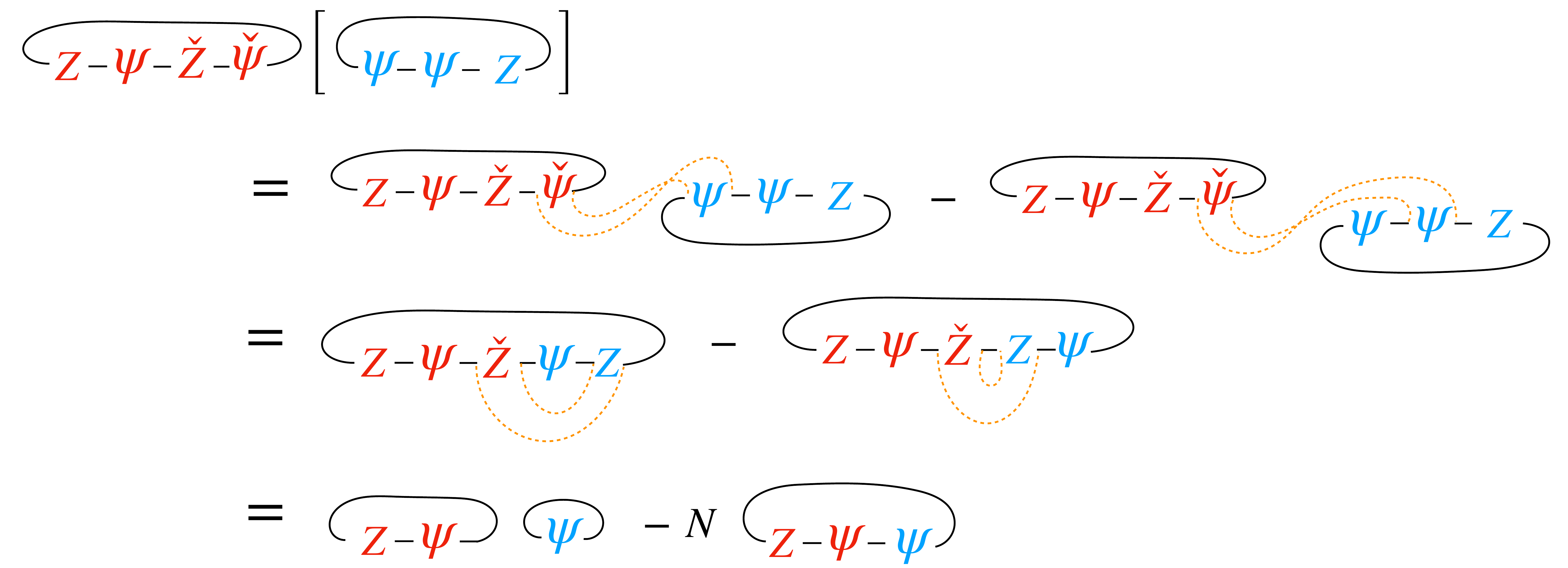}
\label{Hexample}
\eeq
where the lines connecting letters represent contraction of gauge indices within a trace and the orange dashed lines are from the propagator. The letters in red and blue are from the Hamiltonian and the state respectively. The minus sign in the second line comes from moving $\check{\psi}$ across $\psi$ to act on the second letter.

The action of the Hamiltonian is the same for both $SU(N)$ and $U(N)$. (The trace term cancels in the commutator. See \cite{beisert} for a discussion.) In the $SU(N)$ theory, the trace of any single letter is zero so the RHS of (\ref{Hexample}) ends up being $-N\tr(\psi\psi Z)$. Similarly, adding up the contributions from all the terms of $H$, we get
\beq
H \cdot \tr\(\psi\psi Z\)=12\tr\(\psi\psi Z\) \, .
\eeq
Not only is the action on this trace very simple, but it actually yields back the same state in this example with the famous energy $12$ which keeps showing up in this note. That is because $\tr(\psi\psi Z)$ is yet another manifestation of the famous Konishi multiplet, this time in the form of the so-called $SU(1|1)$ Konishi representative. 

\section{$SU(2)$ Details} \la{su2ap}
The $SU(2)$ NHW basis comprises two single traces, three double traces and one triple trace: 
\beqa
&&\!\!\!\!\!\!\mathbb{v}_1    = \tr (Z^3X^3)-\tr(ZX)^3, \nn\\
&&\!\!\!\!\!\!\mathbb{v}_2    =\tr (Z^2XZX^2)+\tr (Z^2X^2ZX)-2 \tr(ZX)^3, \nn\\
&&\!\!\!\!\!\!\mathbb{v}_3    =\tr (X^3) \tr (Z^3)-\tr( ZX^2) \tr (Z^2X), \nn\\
  &&\!\!\!\!\!\!  \mathbb{v}_4 =\tr (Z^2) \tr (ZX^3)+\tr (X^2) \tr( Z^3X)-\nn\\
  &&
\qquad\qquad\qquad\qquad -2 \tr(ZX) \tr(ZXZX), \nn\\
    &&\!\!\!\!\!\!\mathbb{v}_5 =\tr(ZX) \tr (Z^2X^2)-\tr(ZX) \tr(ZXZX),\nn\\
    &&\!\!\!\!\!\!\mathbb{v}_6 =\tr (X^2) \tr (Z^2) \tr(ZX)-(\tr (ZX))^3\,. \la{NHWbasis}
\eeqa
Let us check that one of its elements -- the first say -- is indeed killed by acting twice with the raising operator: 
\beqa
\!\!\!&\mathbb{v}_1   \!\!\!\! &={\color{red} 1}\tr (ZZZXXX)\!-\!{\color{blue} 1}\tr(ZXZXZX) \nn\\ 
\!\!\!&(J_+)\cdot\mathbb{v}_1 \!\!\!\!&={\color{red} 2}  \tr (ZZZZXX) \!+\! ({\color{red} 1}-{\color{blue} 3})\tr(ZZZXZX)\nn \\
\!\!\!&(J_+)^2\cdot\mathbb{v}_1 \!\!\!\!&=2\!\times\! 2 \tr (ZZZZZX) \!+\! (1-3)\!\times\! 2\tr(ZZZZZX) \nn \\
\!\!&&=0\,. \nn
\eeqa
Note that we did not include in the NHW basis (\ref{NHWbasis}) the highest-weight states which would be killed by a single action of the raising operator as those would form a closed block by themselves. In other words, the NHW basis (\ref{NHWbasis}) is made of those operators who are one unit away from being HW. 

In this sector, the Hamiltonian takes the simple form
\beq
H=\frac{-2}N\tr[Z,X][\check{Z},\check{X}] \, .
\eeq
At large $N$, the strings become free and states with different numbers of traces do not mix. The eigenstates in the NHW basis at large $N$ are as follows. There are two massive single traces,
\beq
|\psi\>^{N=\infty}_{\gamma=10\pm 2\sqrt{5}}=
2\mathbb{v}_2-(3\mp\sqrt{5})\mathbb{v}_1  \nn
\eeq
and a double trace made out of a Konishi $SU(2)$ representative dressed by a $1/2$-BPS 20' operator (a single graviton) 
\beqa
|\psi\>^{N=\infty}_{\gamma=12}= \mathbb{v}_5=|\texttt{graviton}\rangle \times |\texttt{su(2) konishi}\rangle \label{konishiSU2ap} \, .
\eeqa
The following states at large $N$ are the multigravitons,
\beqa
|\psi\>^{N=\infty}_{\gamma=0}=\begin{cases}
    \mathbb{v}_3&\\
    \mathbb{v}_6&\\
    -3\mathbb{v}_4+4\mathbb{v}_5&
\end{cases} \, .
\eeqa
It is instructive to recall how the Wick contraction matrix is computed. Consider $\mathbb{W}_{11}$ given by the wick contraction of two $\mathbb{v}_1$'s. At leading order at large $N$ we have that the Wick contraction matrix will be dominated by the planar contributions $\mathbb{W}_{11} \simeq \< \mathbb{v}_1\mathbb{v}_1 \>_\texttt{planar}$ given by
\beq
\includegraphics[scale=0.15]{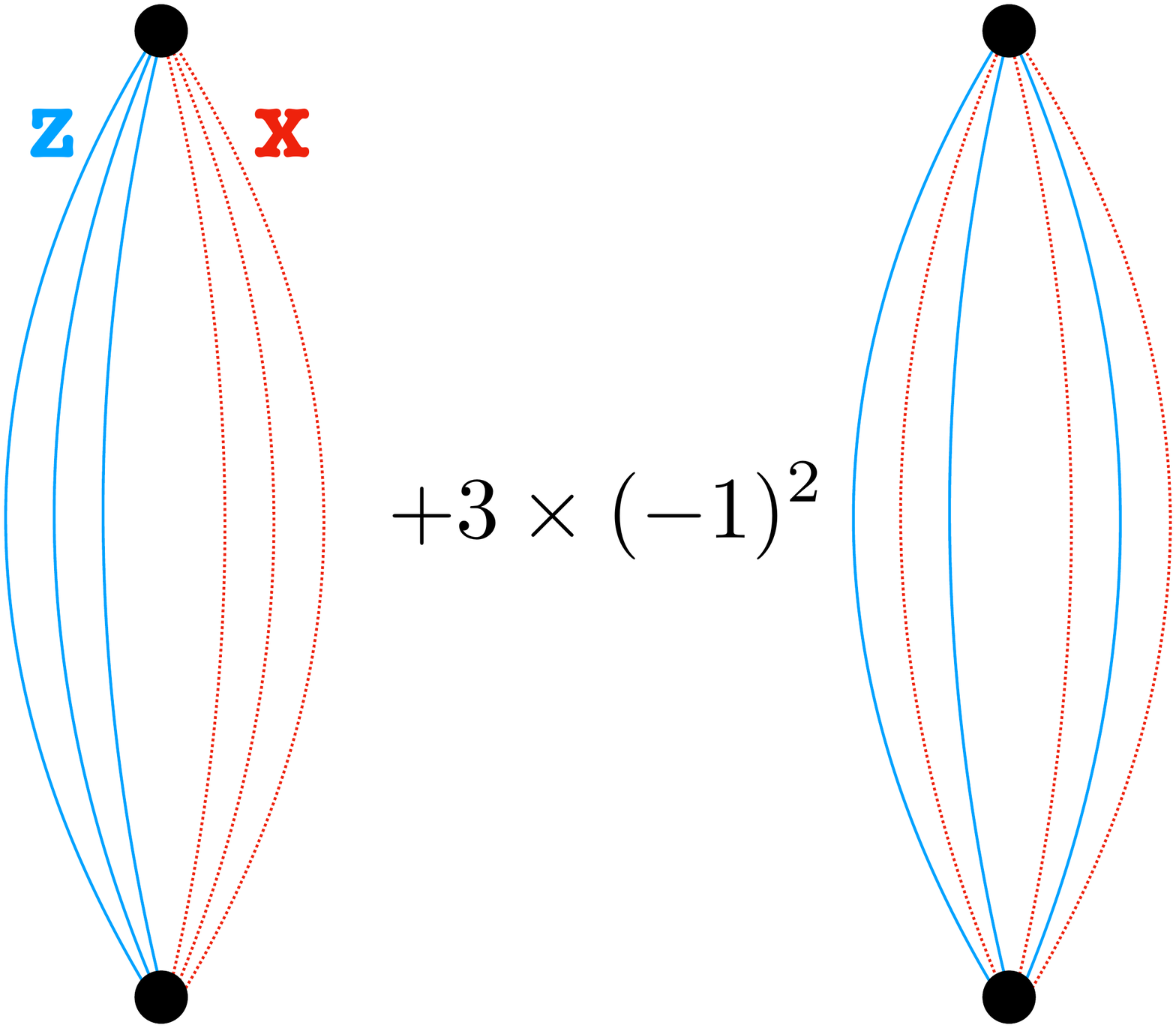}
\eeq
We can replace each propagator here by the $SU(N)$ propagator given by the right hand side of (\ref{propagator}), 
\beq
\includegraphics[scale=0.12]{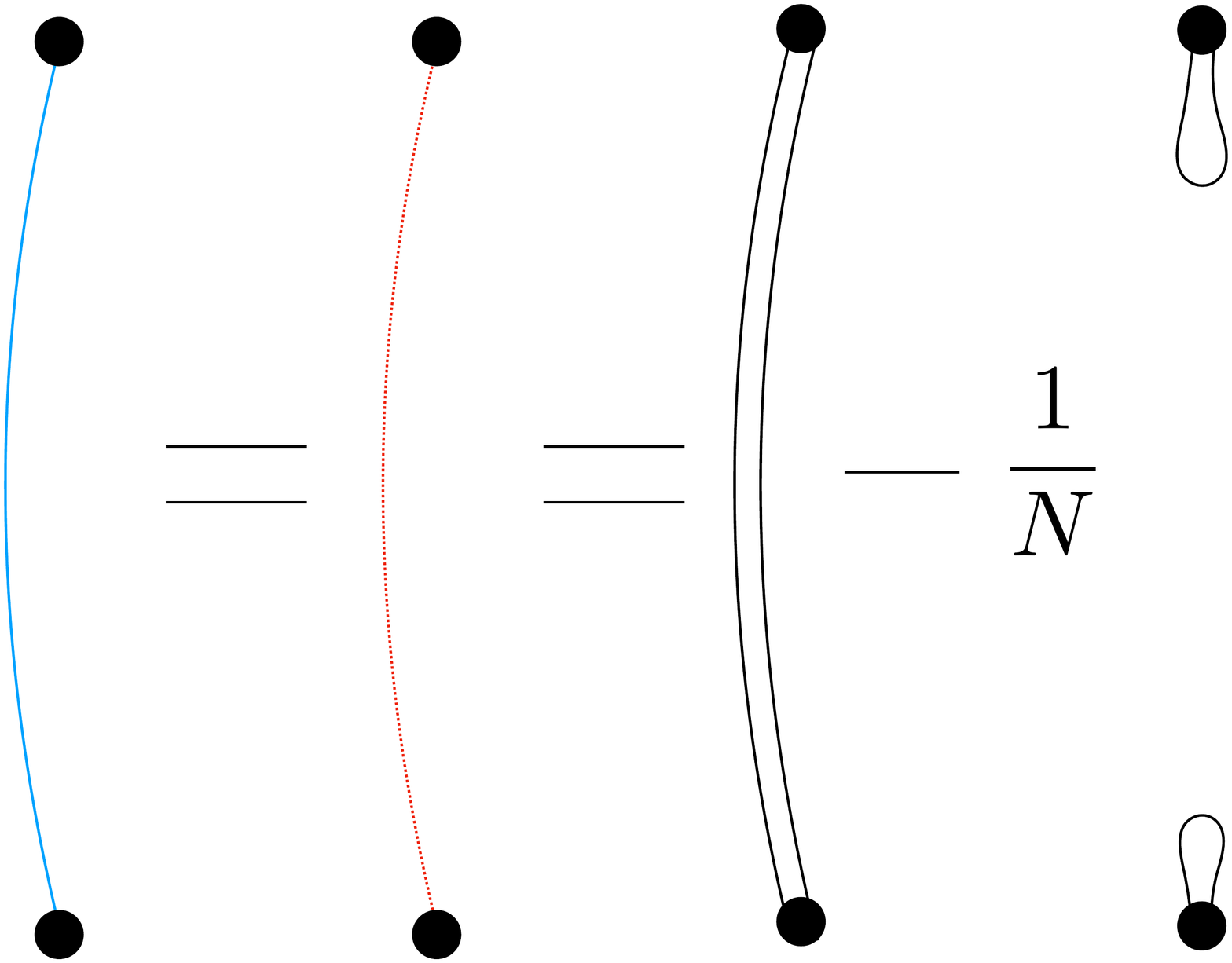}
\eeq
to get 
\beq
\includegraphics[scale=0.15]{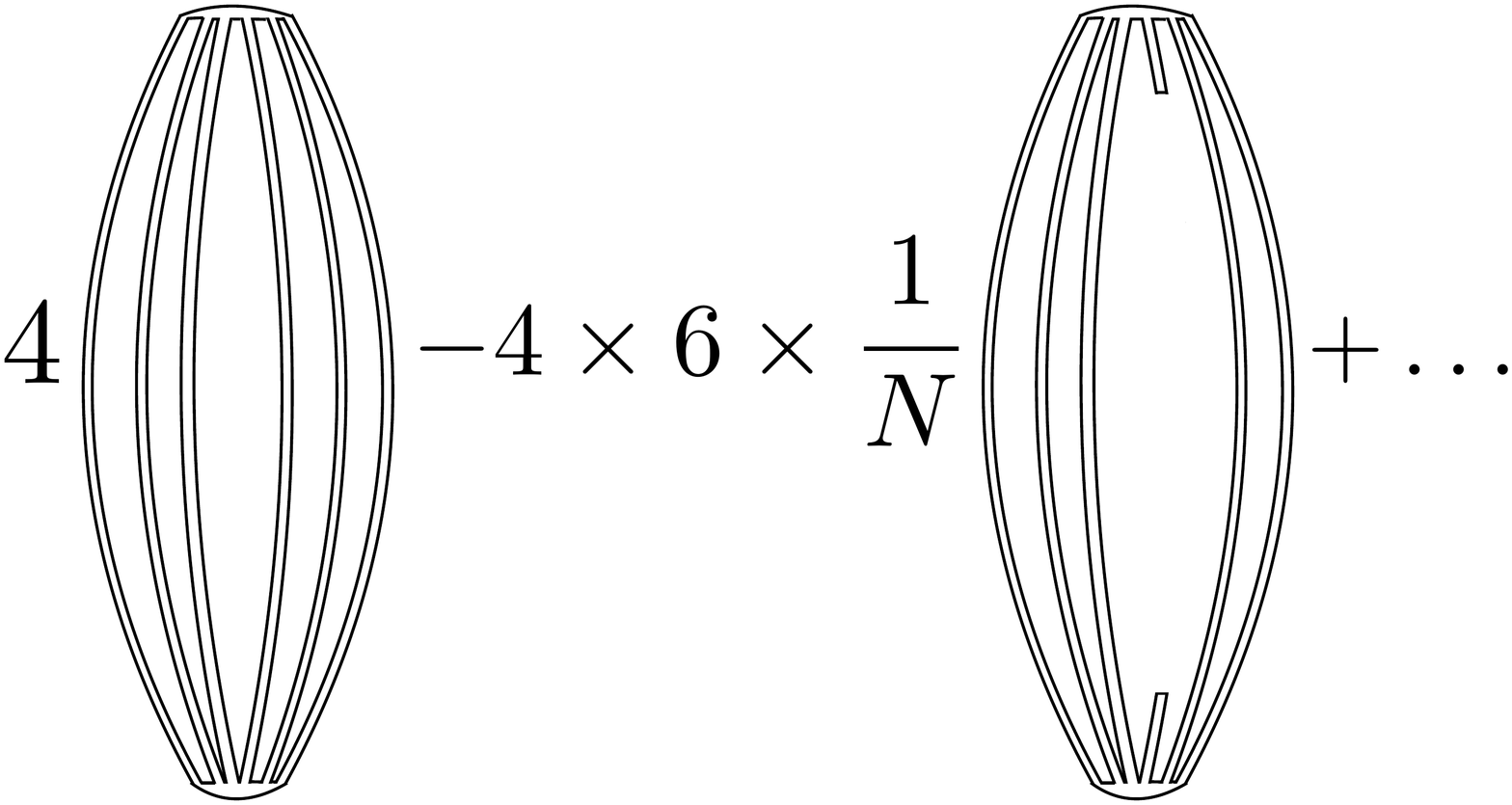}
\eeq
where we used the second term of the propagator once; using it more times leads to more suppressed terms in $1/N$ inside the dots. Once we transformed Wick contractions into pictures like this all we need to do is assign a factor of $N$ for each colour loop.\footnote{Note that at this point -- when we reached the point ot evaluating loops assigning them a weight $N$ -- it is not a big deal to declare this weight to be non-integer.} The first term has six loops and is thus leading while the second term has five loops; all in all we see that 
\beq
\< \mathbb{v}_1\mathbb{v}_1 \>_\texttt{planar}= 4  N^6-2 4 N^4+O(N^2)\,.
\eeq
Note that the leading term perfectly matches the leading term in the first element of the Wick contraction matrix $\mathbb{W}$ in (\ref{Wsu2}); to match the other ones we would need to add to this computation the other non-planar Wick-contractions as well which kick in at order~$N^4$. 



\section{Level Crossing Lift at Two-Loops} \label{crossingSU2}
Sometimes levels still cross at finite $N$ as we can see in figure (\ref{bigFig}). This is an accidental remnant of integrability, and at higher loops we expect all such crossings to be resolved. As illustration we consider here a simple $SU(2)$ sector example where levels cross at one loop but such crossing is then nicely lifted at two loops. 

Consider the  subspace in the $SU(2)$ sector that 
\begin{itemize}
    \item is made out 5 $Z$'s and 3 $X's$, 
    \item is NHW, i.e., killed by $J_+^2$ but not by $J_+$, and,
    \item has even parity.
\end{itemize}
There are nine states in this invariant subspace. As before, we can compute the one-loop anomalous dimensions of these states using (\ref{hamiltonian}). They are the gray lines of figure~\ref{fig:2loop}. The third and fourth energy levels cross at $N\sim 4.3$ as clearly seen in the inset. 

The dilatation operator in the $SU(2)$ sector is known at two-loop order \cite{beisertStaudacherKristjansen,Kim:2003rza}. It takes the following form,
\beqa
H_2&=&\frac4N :\tr[Z,X][\check{Z},\check{X}]: + \nn \\
&&\frac2{N^2}:\tr[Z,X][\check{Z},[Z,[\check{Z},\check{X}]]]:+\nn\\
&& \frac2{N^2}:\tr[Z,X][\check{X},[X,[\check{Z},\check{X}]]]:
\eeqa
Now, diagonalizing the operator $H+\epsilon H_2$, we find the corrected spectrum, shown in figure \ref{fig:2loop} for a small $\epsilon=0.01$ in orange. As seen in the inset, the levels no longer cross but are instead repelled. 

The levels crossings in figure \ref{bigFig} are also expected to be resolved at higher loops in a similar fashion. However, we cannot check this in a straightforward manner because the two-loop non-planar Hamiltonian for the general psu$(2,2|4)$ sector is not known.

Finally, note that the dilatation operator $\mathbb{H}$ could in principle become non-diagonalizable at non-integer N when there are coincident eigenvalues i.e. level-crossings. Since we expect accidental crossings to be resolved at higher loops as illustrated here, we do not worry about this possibility.


\begin{figure}
    \centering
    \includegraphics[width=\linewidth]{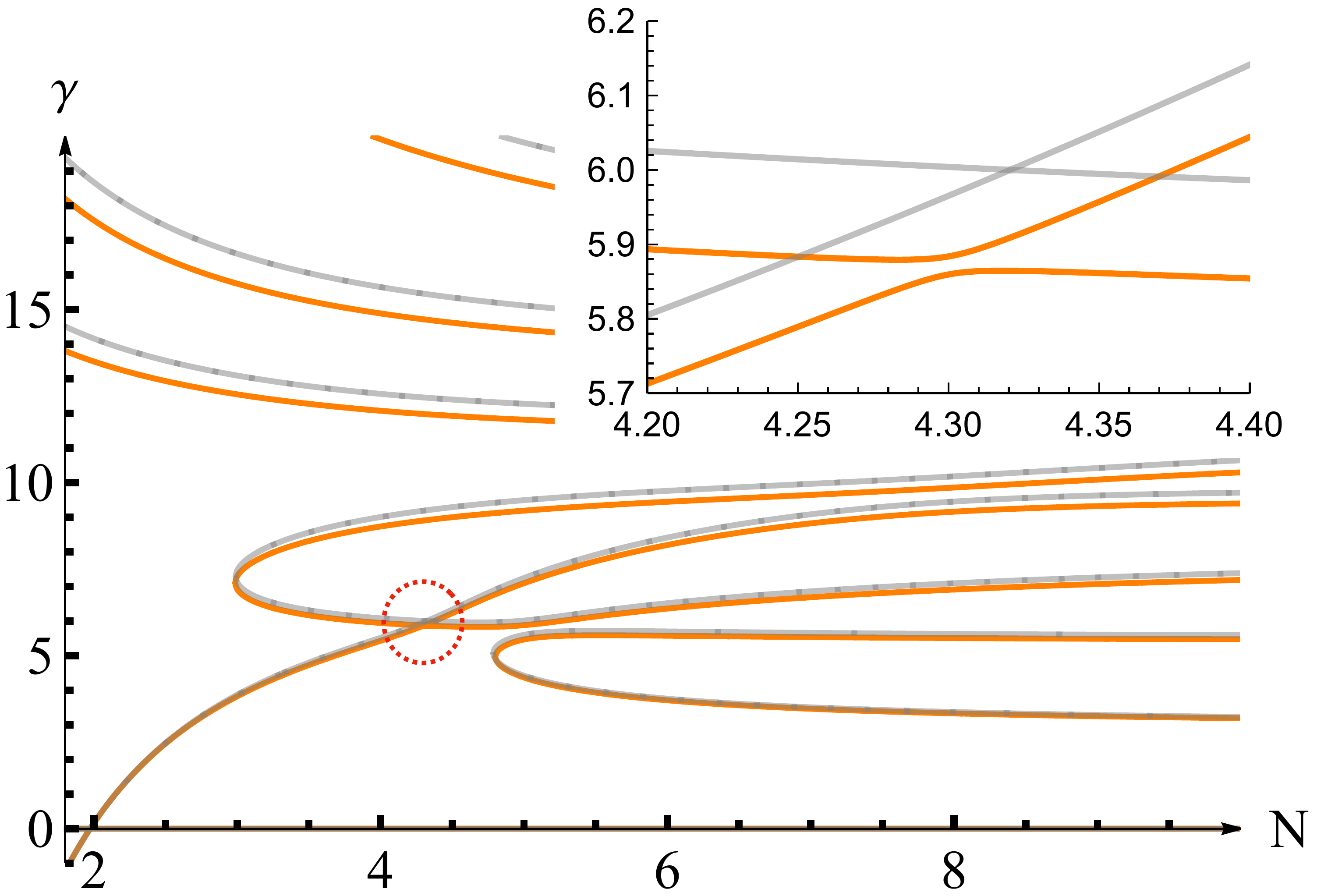}
    \caption{The gray lines are the one-loop anomalous dimensions and the orange lines are the eigenvalues of $H+\epsilon H_2$ for $\epsilon=0.01$. There is an accidental one-loop level crossing, highlighted by the dashed red circle. At two-loops, the levels are instead repelled as seen in the inset.}
    \label{fig:2loop}
\end{figure}

\section{Shortening Conditions} \la{sec:rec}
We review the short unitary representations of the $\text{su}(2,2|4)$ superconformal algebra \cite{cordova2019multiplets, Dolan:2002zh}.

When the classical charges of the superconformal primary (SCP) satisfy one of the so-called shortening conditions, certain state in the multiplet becomes null and the multiplet is cut short. In su$(2,2|4)$, there are three types of short multiplets w.r.t $Q$ denoted by $A_1, A_2$ and $B_1$. The long multiplet is denoted by $L$. Analogously, under the action of $\tilde Q^{\dot\alpha}$, the multiplets are labelled by $\bar L,\bar A_1,\bar A_2$ and $\bar B_1$. The shortening conditions and the corresponding null states are listed in the table \ref{tab:short}.

\begin{table}[h]
\begin{tabular}{l|l|l|l|}
\cline{2-4}
                            & \multicolumn{1}{c|}{SCP}                                                  & \multicolumn{1}{c|}{UB $\Delta=...$} & \multicolumn{1}{c|}{$Q$ Null State}                       \\ \hline
\multicolumn{1}{|l|}{$A_1$} & $[J,\bar J]^{(R_1,R_2,R_3)}_{\Delta}$ & $2\!+\!2J\!+\!\sum\limits_i R_i$            & $[J\!-\!\tfrac{1}{2},\bar J]^{(R_1+\tfrac{1}{2},R_2+\tfrac{1}{2},R_3+\tfrac{1}{2})}_{\Delta+\tfrac{1}{2}}$ \\ \hline
\multicolumn{1}{|l|}{$A_2$} & $[0,\bar J]^{(R_1,R_2,R_3)}_{\Delta}$                    & $2\!+\!\sum\limits_i R_i$            & $[0,\bar J]^{(R_1+1,R_2+1,R_3+1)}_{\Delta+1}$             \\ \hline
\multicolumn{1}{|l|}{$B_1$} & $[0,\bar J]^{(R_1,R_2,R_3)}_{\Delta}$                    & $\sum\limits_i R_i$                 & $[\tfrac{1}{2},\bar J]^{(R_1+\tfrac{1}{2},R_2+\tfrac{1}{2},R_3+\tfrac{1}{2})}_{\Delta+\tfrac{1}{2}}$   \\ \hline
\end{tabular}
\caption{Shortening conditions w.r.t $Q_\alpha\equiv Q^4_\alpha$ reproduced from \cite{cordova2019multiplets} in the notations of Appendix \ref{sec:notations}. There are analogous shortening conditions w.r.t $\tilde Q_{\dot\alpha}$. For the $A_1$ multiplet, in the first line, $J\ge 1/2$. UB stands for Unitarity Bound.}
\label{tab:short}
\end{table}

Existence of the null state implies an existence of a (classically)  $Q_-$-closed state inside the multiplet. For the multiplets of type $A_1$, the null state is
\begin{align}
Q_{\alpha} \ket{\text{SCP}}^{\alpha} = Q_+ \ket{\text{SCP}}_- - Q_- \ket{\text{SCP}}_+ \, .
\end{align}
From this, it follows that the following state must also be null:
\begin{align}
Q_- Q_+ \ket{\text{SCP}}_+ &= - Q_+ Q_- \ket{\text{SCP}}_+ \\
&= -Q_+\( Q_+\ket{\text{SCP}}_- -  Q_{\alpha}\ket{\text{SCP}}^\alpha \)\\
&= Q_+ Q_{\alpha}\ket{\text{SCP}}^{\alpha} \, .
\end{align}
Therefore, $Q_+\ket{\text{SCP}}_+$ is $Q_-$-closed. The BH states (\ref{quantumNumbers}) are of this form.

One can check that the multiplets of type $A_2$ also contain a $Q_-$-closed state, which is again $Q_+\ket{\text{SCP}}$. The SCP of the multiplet $B_1$ is itself $Q_-$-closed.

Whether the operators above remain $Q_-$-closed at quantum level, depends on whether the short multiplet recombines into a long multiplet and obtains an anomalous dimension.

\section{More Chaos, More ``Black Holes"}
In this appendix we want to find some simple evidence for the following:
\begin{itemize}
\item The so-called BH state -- the state whose energy vanishes at $N=2$ -- is pretty typical; the other states with non-vanishing energies behave similarly. They are not any \textit{less} black-hole-like. 
\item MG states, however, exhibit a somewhat different qualitative behavior. They have bigger disintegration probabilities in line with the intuition that they are dual to gases of loosely connected gravitons of sorts, rather than compact BH-like objects. 
\item At large $N$ chaos decreases in agreement with the expectation that -- at fixed quantum numbers -- we will be describing a bunch of free strings governed by integrability once $N\to \infty$.
\end{itemize}
Of course, this is all quite qualitative -- we are at weak coupling after all -- and will be based on some simple statistical analysis of the various states obtained when diagonalizing the non-planar dilatation operator for states with the BH and MG small quantum numbers.
\begin{figure}[t]
    \centering
    \includegraphics[width=0.5\textwidth]{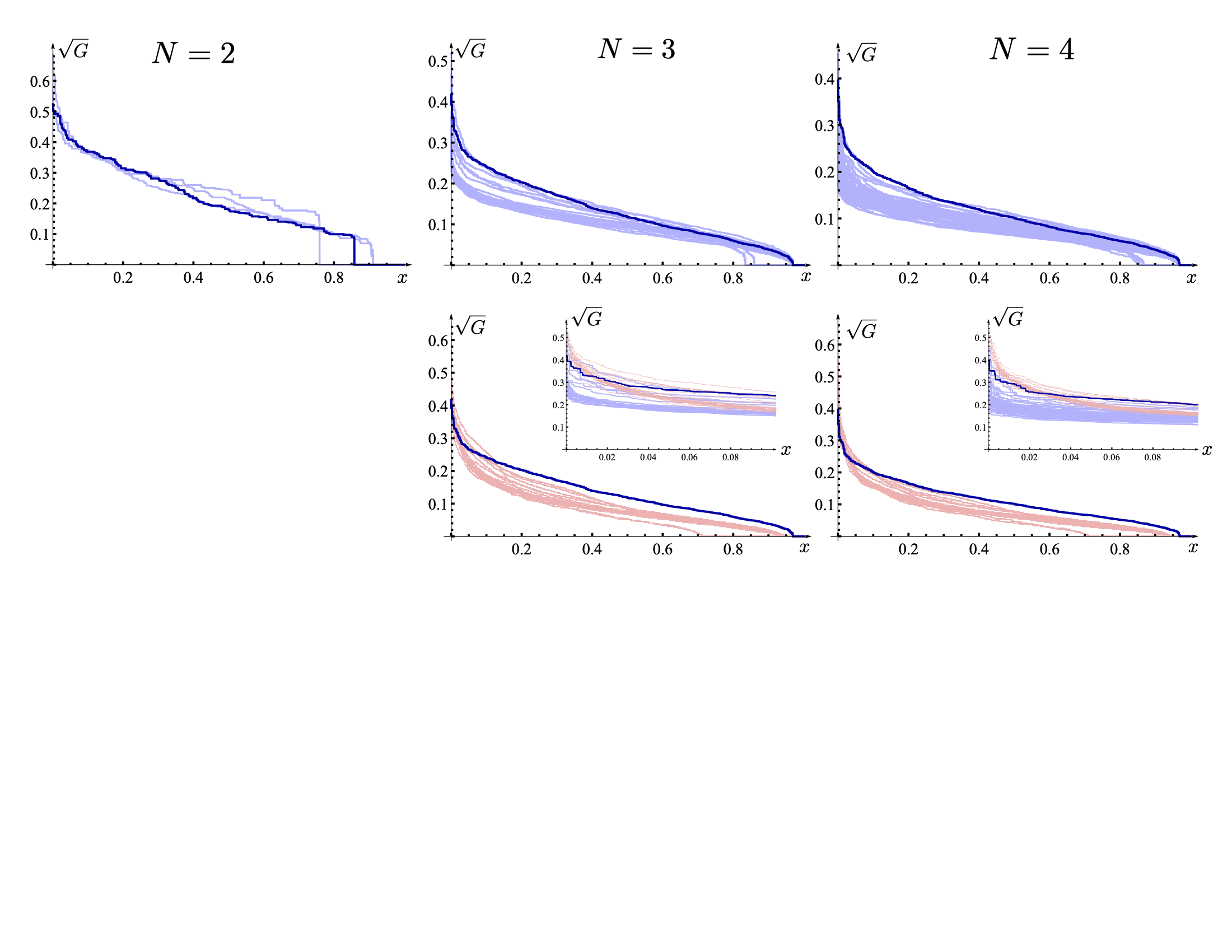}
    \vspace{-2.7cm}
    \caption{Correlators. BH's in the top, MG's in the bottom. MG have bigger disintegration $G$'s small $x$ region where these are bigger; otherwise the BH state is always (one of the) state(s) which disintegrates easier, i.e. with larger $G$.}
    \label{correlators}
\end{figure}

The first evidence for the first two bullet points is represented in figure \ref{correlators}. On the top row of this plot we are depicting thousands of correlation functions per each of the $69$ states with the black hole quantum numbers while in the bottom row are are plotting the same correlation functions involving instead the $27$ multi-graviton states with that field content.\footnote{Although they have the same fields, they do \textit{not} have the same quantum numbers. The 69 states in blue are the so-called BH sector described by all singlet operators with even parity that are primaries under psu$(1,2|3)$ as described in the main text. In particular, there are no multi-graviton states once we impose this. The 27 pink states, instead, are defined as the states with the same field content and with zero energy, independently of any other quantum number restriction.} With the field content of the BH state there are 8013 states. Out of these, 4893 are double or triple traces such as  
\beqa
\mathbb{v}_i&=&\tr(\psi^3 \phi^{23} )\times \tr (\phi^{13} \phi^{12}) \times \tr (f f \phi^{12})  \nn 
\\ &\equiv& o_1(x)\times o_2(x) \times o_3(x)  \,.\label{exampleBasis}
\eeqa
We can now take one of the $69$ states -- lets call it $O_A$ -- and study it disintegration into the constituents of these multi-traces. I.e. we take any $m$-trace basis element such as (\ref{exampleBasis}) and split it into $m$ single trace operators at different location to define
\beq
\mathbb{G}^A_i \equiv \< O_A(0) o_1(x_2)\dots o_m(x_m) \>/(\mathcal{N}_A n_1 \dots n_m) \la{Gi}
\eeq
where $\mathcal{N}_A$ and $n_i$ are normalizations given by the operator two point functions. Finally, this quantity is not a number, it depends on the various locations $x_j$. To make it a number we assume these locations are the vertices of the basis of a regular pyramid whose tip is the origin and where all the vertices are one unit away from the tip so that the various propagators are just $1$.\footnote{The field strength $f$ propagator is $2$ because of the derivatives.} This is a very convenient definition because it is such that the extremal correlator (\ref{Gi}) is directly computed by the $8013\times 8013$ Wick contraction matrix $\mathbb{W}$ defined in (\ref{innerProduct}) as 
\beq
G_i^A = \frac{(\mathbb{W}\cdot \psi^{(A)})_i}{ (\psi^{(A)}\cdot \mathbb{W}\cdot \psi^{(A)})^{1/2} n_1\dots n_m  }
\eeq
where $O_A=\sum_{j} \psi_j^{(A)} \mathbb{v}_j$. The normalizations $n_1 \dots n_m$ are computed similarly through Wick contractions but those are much simpler to compute since these are very small operators. For each $A=1\dots A_\text{max}$ we make a list of $G_i^A$ with $i$ ranging over the $i_\text{max}$ values corresponding to the basis elements which are either double or triple traces. Then we take the square root of the absolute value of each element and sort the list. We then define $x\equiv i/i_{max} \in [0,1]$ and plot the outcome. This is how we got the many blue lines in figure \ref{correlators}. For $N$ large enough there are $A_{max}=69$ blue curves each with $i_{max}=4893$ points but for low $N$ we have trace relations so that we will have less curves with less points. We can repeat this exercise for $O^A$ replaced by any one of the multigraviton states $O^A_\text{MG}$. This is how we got the red curves in the figure. As we can see in the figure, these correlation function show a nearly universal behavior and we expect deviations from typicality to be exponentially suppressed in the entropy \cite{Balasubramanian:2005mg}.

We like to think of (\ref{Gi}) as a sort of decay process where the operator $O^A$ disintegrates into a bunch of $o_1 \dots o_m$. We further like to imagine that this quantity being small is a sort of signal that the object is stable and does not want to be broken apart. Of all the blue curves, the BH state is, in this sense, among the most decay-prone ones at least for $N\ge 3$ where it clearly is close to the top of the pack of all states with those quantum numbers. Perhaps we should think of all these states as black holes of sorts and the one which becomes $1/16$-th BPS one at $N=2$ is in a way the least tight of all. 

\begin{figure}[t]
    \centering
    \includegraphics[scale=0.3]{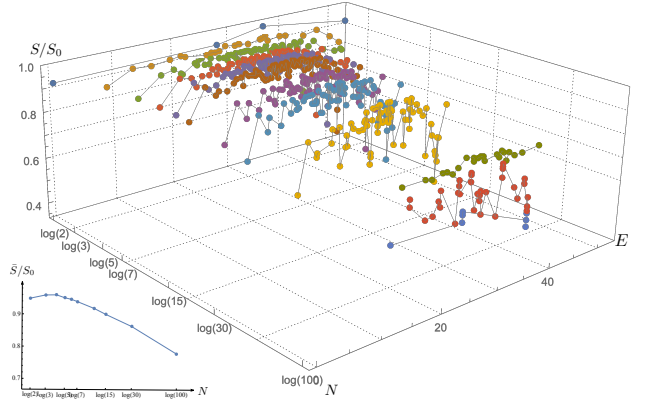}
    \vspace{-1cm}
    \caption{Eigenstates' information entropies. At large $N$  (illustrated here with $N=100$) integrability kicks in and we see a big spread in the entropies indicating a sort of chaos breakdown; we also see that single (green), double (red) and triple (purple) traces start to cluster (which is why at that large $N$ only did we colour them differently). In the lower inset we averaged over the entropy in each $N$ cloud.}
    \label{figchaos}
\end{figure}

Multi-gravitons behave slightly differently it seems, for small $x$ corresponding to the most significant decay channels. As illustrated in the second row -- most notably in the insets -- for small values of $x$ they produce bigger decays $G$ compared to all blue states. Perhaps this is in line with the expectation that these should be some sort of graviton gases evoking a picture of a bunch of loosely right elements which one can easily tear apart?

Finally we went on to analyse the inner structure of the various $O_A$ states using the so-called information entropy content of the states, a nice quantity we learned from~\cite{TristanEtAl}, a reference that we follow closely in the rest of this appendix. For each state $O_A$ we define a vector 
\beq
\mathbb{w}_A \equiv  \sum_{i} \psi_i^{(A)} \sqrt{\mathbb{W}}\cdot\mathbb{v}_i \equiv \sum_i c_i \mathbb{v}_i  \la{wA}
\eeq
Then the norm of the state is simply the usual scalar product, $\mathcal{N}_A=(\mathbb{w}_A \cdot \mathbb{w}_A)^{1/2} = \sqrt{\sum_i |c_i|^2}$.\footnote{The advantage of this definition is to absorb the Wick contraction matrix into the states rendering their scalar product orthonormal in contradistinction to the multi-trace basis. While was not done in \cite{TristanEtAl}, it was highlighted there that this would be safer to make the scalar product more canonical. We followed their advice in this section using the $\sqrt{\mathbb{W}}$ matrix. We also tried doing the more straightforward computation using the multi-trace basis -- computationally much simpler -- and the conclusions are basically the same. }   Here the square root of the (real and symmetric) Wick contraction matrix is the matrix such that 
\beq
\sqrt{\mathbb{W}} \cdot \sqrt{\mathbb{W}} = \mathbb{W} 
\eeq
which we can compute. (It is not completely trivial, it takes a few minutes for each $N$ choice with \texttt{Mathematica} using \texttt{MatrixPower[W,1/2]}.)

The information entropy is then defined as 
\beq
S\equiv -\sum_i \frac{|c_i|^2}{\mathcal{N}_A} \log \frac{|c_i|^2}{\mathcal{N}_A} \la{Sdef}
\eeq
and the prediction is that a chaotic system should have an entropy consistent with Gaussian orthogonal ensemble random matrix theory. It should be well approximated by 
\beq
S_0 \equiv \log(I)+\log(2)+\gamma_E-2
\eeq
where $I$ is the number of basis elements appearing in the sums (\ref{wA}) and (\ref{Sdef}) and $\gamma_E$ is the Euler's constant. Indeed, if we plot $S/S_0$ for the various states at various values of $N$ we see that the ratio $S/S_0$ is indeed very close to $1$ as depicted in figure \ref{figchaos}. In the figure's inset we plotted, for each $N$ the average of $S/S_0$ over all states with the BH quantum numbers. We see that as we increase $N$ the system starts deviating from random matrix theory universality. This is expected as integrability should kick in. Indeed, at $N=100$ we see already that the ratio is around $0.7$ and -- moreoever -- the spread around this average is more pronounced with single, double and triple traces clustering around three separated clouds. 

Physically this makes sense. If $N$ is very large compared to the various charges of the states, we are describing a bunch of free strings so their dynamics should indeed decouple into groups (depending on the number of strings) and each group should be simply described by order (integrability) rather than chaos.

\section{Adiabatic Toy Model} \label{adiabatic}
\begin{figure}[t]
    \centering
    \includegraphics[scale=0.7]{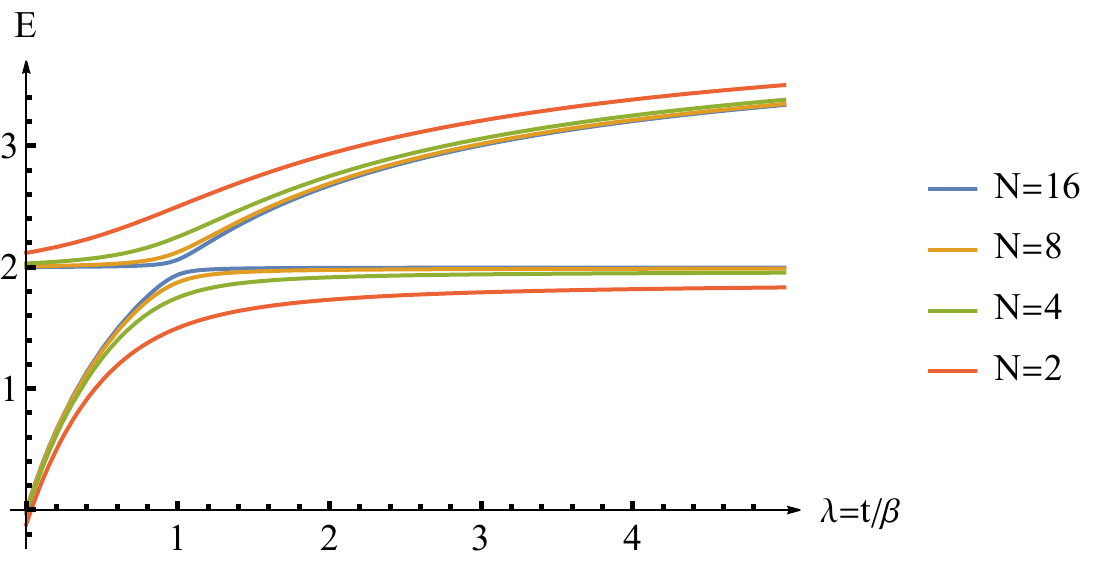}
    \caption{Eigenvalues of $\mathbb{H}(\lambda)$. The level splitting is smallest when $N$ is largest.}
    \label{evToy}
\end{figure}

Consider the toy Hamiltonian 
\beq
\mathbb{H}(\lambda)=\left(
    \arraycolsep=1pt\def\arraystretch{1.5}
\begin{array}{cc}
 2 & \frac1N \\
 \frac1N & \frac{4\lambda}{1+\lambda}\\
\end{array}
\right) \label{Htoy}
\eeq
whose eigenvalues as function of $\lambda$ are plotted in figure \ref{evToy}. We see that $N$ controls how violent the level repulsion is: For large $N$ the levels only repel when they are extremely close to each other while for small $N$ the level repulsion is quite smooth.

We can now consider an experimental setup where $\lambda$ is allowed to grow with time as $\lambda=\beta t$ where $\beta$ is a positive constant that governs the growth rate. Smaller values of $\beta$ correspond to a slower ramp-up. We want to consider the time dependent Schrodinger equation 
\beq
\frac{1}{i}\left(\begin{array}{c}
\psi_1'(t)\\
\psi_2'(t)
\end{array}
\right)= \mathbb{H}(\lambda(t))\cdot \left(\begin{array}{c}
\psi_1(t)\\
\psi_2(t)
\end{array}
\right)
\eeq
(Note that as $t \to \infty$ the Hamiltonian asymptotes to a time independent Hamiltonian.) This is a nice toy model since this equation can be solved analytically.

\begin{figure}[t]
    \centering
    \subfloat[\label{(a) N=16}]{
        \includegraphics[width=0.5\textwidth]{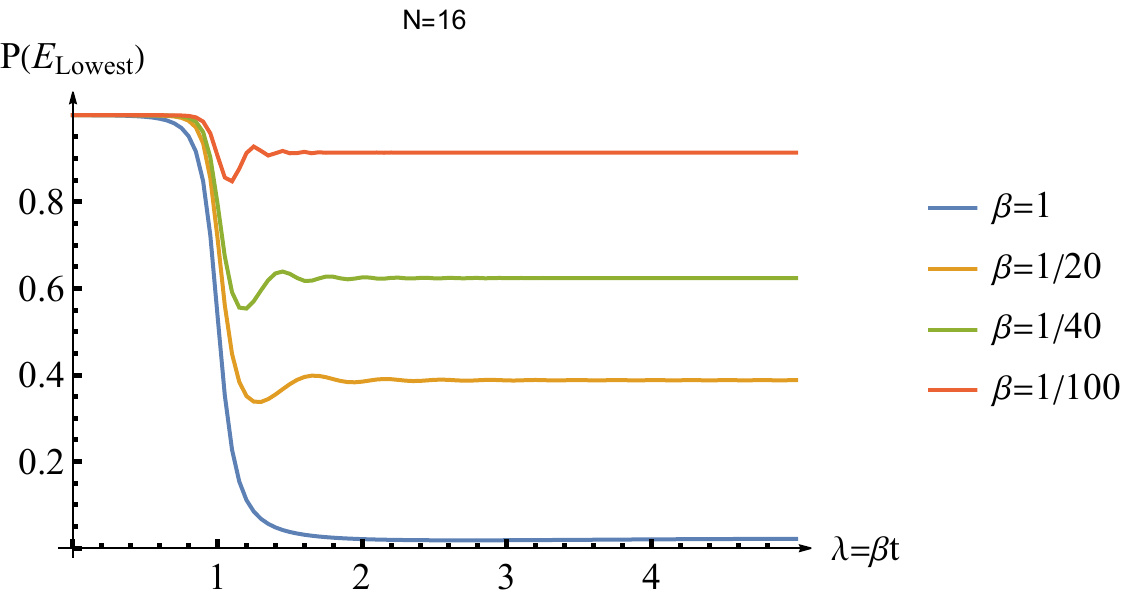}}
    \hfill
    \subfloat[\label{(b) N=2}]{
        \includegraphics[width=0.5\textwidth]{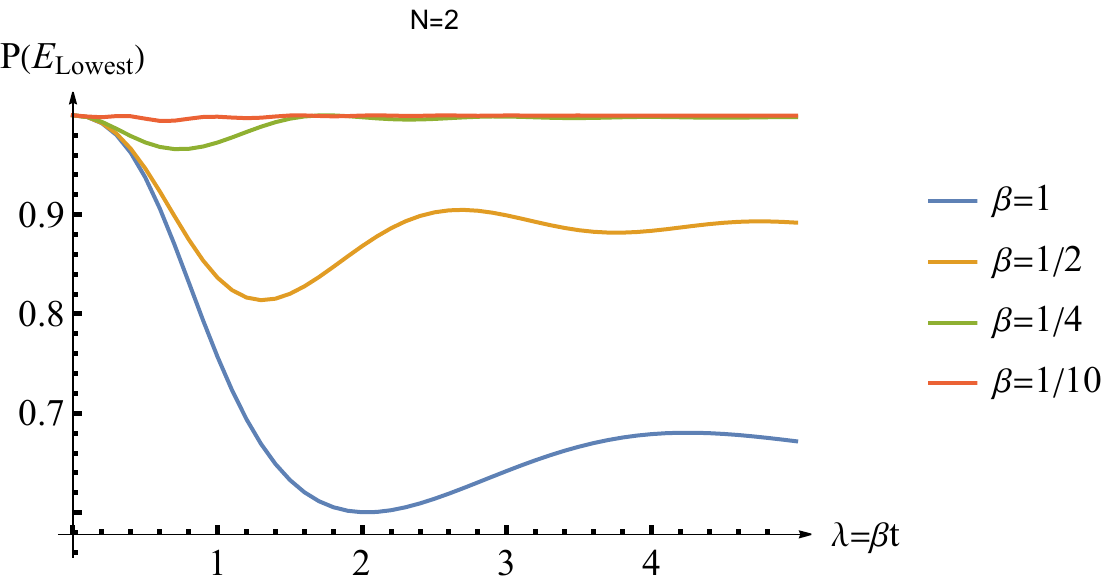}}
    \hfill
    \caption{Probability of staying in the lowest eigenvector of $\mathbb{H}(\lambda)$ when coupling reaches $\lambda=\beta t$ starting from the lowest eigenvector at $\lambda=0$. }
    \label{probToy}
\end{figure}

We can start as the lowest energy eigenvector of $\mathbb{H}(0)$ as initial conditions and compute the probability of being in the lowest energy eigenvalue of $\mathbb{H}(\lambda)$ at any $t>0$, see figure \ref{probToy}. We see that at $N=16$, we need very small $\beta\sim 0.01$ in order to track the lowest energy state with high probability. On the other hand for the samller $N=2$, the probability is already very high at $\beta=0.25$. The larger the $N$, the slower we need to change $\lambda$ in order to track the lowest state. This makes sense because at large $N$ the levels get much closer which facilitates an easier crossing to the excited level.


If experimentalists were to continue the Konishi-plus-two-gravitons state (\ref{konishiBH}) from weak to strong coupling at large $N$ in a similar fashion, they would encounter such level repulsions. Hasty experimentalists \textit{moving} too fast will jump from level to level and follow the integrability prediction and end up with a very massive string state at strong coupling. Patient experimentalists moving very slowly (the larger $N$ the slower we would need to move) would stay in the lowest energy state and end up with a bunch of gravitons at strong coupling. These patient experimentalists would be following the \textit{true} state as in this toy model. 

\section{The State in Two Pictures}\label{entanglementAp}
 Written down in components (i.e. multi-traces) the states we are dealing why are intimidating -- see figure \ref{hugeFig}. The large $N$ state with about 500 terms can be simplified to (\ref{konishiBH}) using the entanglement operator while the $N=2$ state with its almost 8000 terms becomes compactified into the three-liner (\ref{N2state}) once we take advantage of trace relations as well as the susy operator. 

\begin{figure*}[t]
   \includegraphics[width=0.85\textwidth]{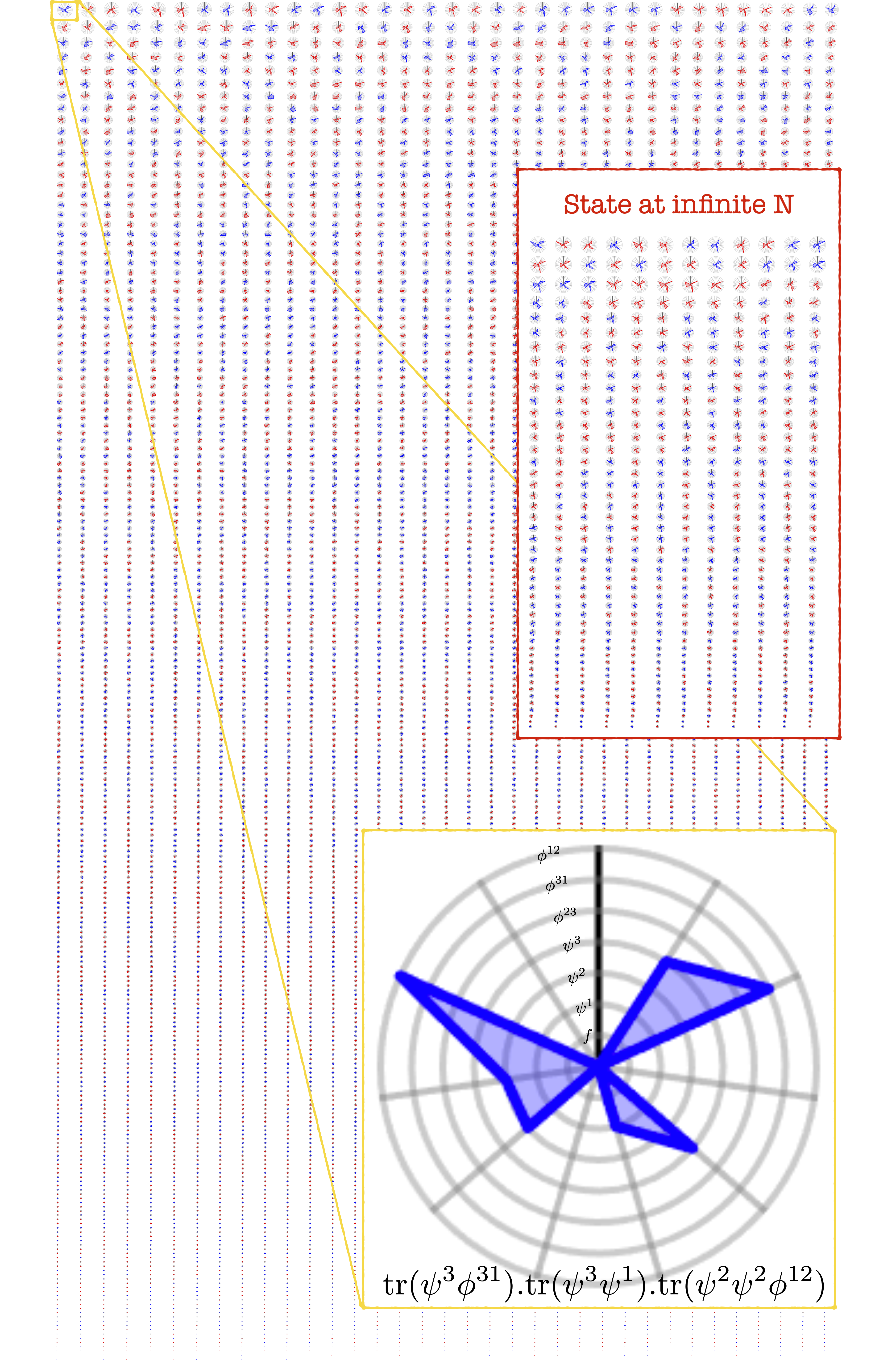}
   \caption{The black hole state for $N=2$ in the trace basis is huge while $N=\infty$ in the red box is much simpler. Each picture is given by a multi-trace following the example in the bottom. Each term is rescaled by square root of the amplitude of the corresponding multi-trace with red and blue indicating the sign of the amplitude. 
   } \label{hugeFig}
 \end{figure*}

\bibliography{BPSBlackHoles} 

\end{document}